\documentclass[aps,prb,twocolumn,groupedaddress,showpacs,amsmath,amssymb]{revtex4}

\usepackage{graphicx}
\usepackage{dcolumn}

\providecommand{\eqref}[1]{(\ref{#1})}

\newcommand{\abs}[1]{\left\vert#1\right\vert}

\newcommand{\jRe}{\Re}

\newcommand{\eps}{\varepsilon}

\newcommand{\phdag}{{\phantom{\dag}}}

\newcommand{\conj}{{*}}
\newcommand{\phconj}{{\phantom{\conj}}}

\newcommand{\jprime}{{\prime}}
\newcommand{\jbracketprime}{{(\prime)}}
\newcommand{\jvecprime}{{\,\jprime}}
\newcommand{\jvecbracketprime}{{\,\jbracketprime}}

\newcommand{\supzero}{{(0)\!}}
\newcommand{\phsupzero}{{\phantom{(}\!}}

\newcommand{\NitrogenGroundStateTermSymbol}{\ensuremath{^1\Sigma_g^+}}
\newcommand{\NitrogenGroundState}{\ensuremath{N_2(\NitrogenGroundStateTermSymbol)}}

\newcommand{\NitrogenDominantMetastableStateTermSymbol}{\ensuremath{^3\Sigma_u^+}}
\newcommand{\NitrogenDominantMetastableState}{\ensuremath{N_2(\NitrogenDominantMetastableStateTermSymbol)}}

\newcommand{\NitrogenNegativeIonResonanceTermSymbol}{\ensuremath{^2\Pi_g}}
\newcommand{\NitrogenNegativeIonResonance}{\ensuremath{N_2^-(\NitrogenNegativeIonResonanceTermSymbol)}}

\newcommand{\PenningME}[1]{V_{\scriptscriptstyle 0#1, \vec{k}}^{\scriptscriptstyle \vec{q}, 1#1}}

\newcommand{\RenormalizedPenningME}[1]{\mathcal{V}_{\scriptscriptstyle 0#1, \vec{k}}^{\scriptscriptstyle \, \vec{q}, 1#1}}
\newcommand{\PenningMEStub}[1]{V_{\scriptscriptstyle\vec{k}\vec{q}}^{\scriptscriptstyle #1}}

\newcommand{\GammaFunction}[1]{\Gamma_{\scriptscriptstyle\vec{k}\vec{q}}\!\left( #1 \right)}

\begin{document}

\title{Auger de-excitation of metastable molecules at metallic surfaces}

\author{Johannes Marbach}
\email[Contact:\,]{marbach@physik.uni-greifswald.de}
\author{Franz Xaver \surname{Bronold}}
\author{Holger Fehske}

\affiliation{Institut f\"ur Physik, Ernst-Moritz-Arndt-Universit\"at Greifswald, 17489 Greifswald, Germany}

\date{\today}

\begin{abstract}
We study secondary electron emission from metallic surfaces due to Auger de-excitation of diatomic metastable 
molecules. Our approach is based on an effective model for the two active electrons involved in the 
process -- a molecular electron described by a linear combination of atomic orbitals when it is 
bound and a two-center Coulomb wave when it is not and a metal electron described by the eigenfunctions of 
a step potential -- and employs Keldysh Green's functions. Solving the Dyson equation for the retarded Green's 
function by exponential resummation we are able to treat time-nonlocal self-energies and to avoid the wide-band 
approximation. 
%Utilizing only the localized character of the bound molecular wave functions and the large value 
%of the molecule's turning point we set up a numerical scheme for the calculation of the secondary
%electron emission coefficient and the spectrum of the emitted electron which is exact within our model. 
Results 
are presented for the de-excitation of \NitrogenDominantMetastableState\ on aluminum and tungsten and discussed 
in view of previous experimental and theoretical investigations. We find quantitative agreement with experimental 
data for tungsten indicating that the effective model captures the physics of the process quite well. For
aluminum we predict secondary electron emission due to Auger de-excitation to be one to two orders of 
magnitude smaller than the one found for resonant charge-transfer and subsequent auto-detachment.
\end{abstract}


% 34.35.+a				Gas-surface interactions
% 68.47.De				Metals => metallic surfaces
% 34.35.+a, 79.20.Rf	Molecule-surface collisions
% 34.50.-s				Molecules => scattering
% 34.35.+a				Molecules => scattering from surfaces
% 79.20.Hx				Secondary electron emission
% 79.20.Fv				Electron emission => Auger emission

\maketitle

\section{Introduction\label{introduction}}

De-excitation of metastable atoms and molecules with simultaneous release of an electron 
is a surface scattering process of great technological importance. De-excitation of atoms 
is used as a surface-sensitive electron spectroscopy~\cite{sesselmann87,harada97,kantorovich00,winter02} 
and de-excitation of molecules is an important process in molecular low-temperature gas discharges. It 
is one of the main wall-based secondary electron emission channels controlling, together with wall recombination
and various volume-based charge production and destruction channels, the overall charge balance in the 
discharge.~\cite{liebermann05} In the de-excitation process both the target and the projectile are composite 
objects. A great variety of reaction channels is thus conceivable making the investigation of this scattering 
process a challenging task, particularly for molecules.

\citeauthor{stracke98}\cite{stracke98} experimentally investigated the de-excitation of metastable 
nitrogen \NitrogenDominantMetastableState\ molecules on a tungsten surface and proposed two main reaction 
channels. Firstly, the Auger de-excitation (also referred to as Penning de-excitation),
\begin{equation}
	\NitrogenDominantMetastableState + e_m \rightarrow \NitrogenGroundState + e_f~,
	\label{penning reaction}
\end{equation}
where $e_m$ and $e_f$ denote an electron inside the metal and a free electron, respectively, and secondly, 
the formation of the \NitrogenNegativeIonResonance\ shape resonance with subsequent auto-detachment,
\begin{equation}
	\NitrogenDominantMetastableState + e_m \rightarrow \NitrogenNegativeIonResonance \rightarrow \NitrogenGroundState + e_f.
	\label{direct reaction}
\end{equation}
\citeauthor{stracke98}\cite{stracke98} concluded that out of these two competing processes 
reaction~\eqref{direct reaction} should be more efficient, as it is a combination of two single-electron 
charge-transfer transitions, whereas~\eqref{penning reaction} represents a less probable two-electron 
transition. Using thermal molecules they measured the energy spectrum of the released electron and estimated 
the overall secondary electron emission coefficient $\gamma_e$, that is, the probability for releasing 
an electron by a single metastable molecule de-excitation at the surface, to be about $10^{-3}-10^{-2}$.
The experimental estimate for $\gamma_e$ does not discriminate between the two reaction channels. It rather 
includes both channels. Indeed, \citeauthor{stracke98}\cite{stracke98} mention that in the spectrum of the 
emitted electron they also observe a weak signal due to Auger de-excitation. It is one order of magnitude 
weaker than the signal due to charge-transfer. 

Based on the assumption that the charge-transfer channel~\eqref{direct reaction} is the dominant 
one \citeauthor{lorente99}~\cite{lorente99} theoretically investigated the de-excitation of 
\NitrogenDominantMetastableState\ molecules on an aluminum surface. The resonance-driven secondary electron 
emission coefficient resulting from their calculated electron emission spectrum is about $10^{-1}$ which is 
one order of magnitude larger than the value \citeauthor{stracke98}~\cite{stracke98} give for tungsten. The 
Penning channel~\eqref{penning reaction} was not considered by Lorente and coworkers.~\cite{lorente99} 
Its strength for an aluminum surface is thus unknown. 

In the present work we adopt the point of view complementary to \citeauthor{lorente99}'s~~\cite{lorente99} 
investigating Penning de-excitation while neglecting any contribution due to resonant charge-transfer. In 
particular for tungsten the efficiency of the Penning process may be comparable to the efficiency of the 
charge-transfer process, because the molecular orbital hosting the hole in the electronic configuration 
of \NitrogenDominantMetastableState\ is roughly $2.5\,eV$ below the bottom of the conduction band of 
tungsten.~\cite{kaldor84,christensen74} To bring this orbital in resonance with conduction band states of 
the metal requires therefore a large image shift and broadening due to the interaction with the metal. 
Rough estimates of these two effects based on what is known about them for alkali atoms interacting with 
surfaces~\cite{gadzuk67} imply that the resonance condition can only be met for vibrationally 
excited \NitrogenDominantMetastableState\ states. Thus, at least for \NitrogenDominantMetastableState\ in 
its vibrational ground state, Penning de-excitation and charge-transfer are eye-to-eye competitors. For 
aluminum the electronic band structure is much more favorable for the charge-transfer 
scenario.~\cite{segall61} Here, the bottom of the conduction band is only $1\,eV$ above the molecular 
orbital in question. This energy difference may be bridged by the combined action of image shift and 
level broadening. 

In order to theoretically analyze the Auger de-excitation of diatomic molecules on metallic surfaces we 
adopt an effective two-electron model, where one electron resides in the conduction band of the metal 
and the other in the excited state of the molecule. The metal is modeled as a half-space containing a 
free electron gas characterized by a work function and a conduction band depth, while the molecule is 
modeled in terms of a two-level system with energy spacing corresponding to the excitation energy of 
the metastable state. Focusing on the situation where the emitted electron (Auger electron) originates
from the molecule we describe it by a two-center Coulomb wave. The coupling between the
molecule and the metal is through the Coulomb interaction between a metal electron and the electron in 
the excited state of the molecule. Image interactions are neglected (except for the emitted electron, 
where we include it in terms of a surface transmission function) because they give rise to a 
hybridization of metallic and molecular single-electron states~\cite{gadzuk67} which we prefer to 
discuss in connection with direct charge-transfer.

Although our model for Auger de-excitation is somewhat crude, it contains
the most relevant degrees of freedom in a reasonable approximation and can be parameterized by energies
which are relatively easy to obtain. In particular the latter aspect is rather important for us, because our 
interest in Auger de-excitation and related processes stems from their relevance for bounded low-temperature 
gas discharges. It is well known that electrons can be released from plasma walls by de-excitation of
metastable species but for most wall materials and projectiles the secondary electron emission coefficient 
is unknown. A flexible, easy-to-use microscopic model for its calculation is thus needed.

Following the lead of Makoshi and coworkers,~\cite{makoshi91,makoshikaji91} who investigated the de-excitation 
of metastable atoms, we employ in the following the Keldysh technique\cite{keldysh65,blandin76,danielewicz84,rammer86} 
to calculate within the trajectory approximation~\cite{hagstrum54} the secondary electron emission coefficient 
and the spectrum of the emitted electron for a diatomic metastable molecule hitting a metallic surface. A 
description of this type of surface collision with Green's 
functions~\cite{makoshi91,makoshikaji91,goldberg92,shao94,dutta01,vicentealvarez98} is mathematically more 
demanding than using rate equations.~\cite{hagstrum54,yoshimori86,penn90,salmi92,Cazalilla98,bonini03} Green's functions 
are however rather flexible in handling the non-adiabaticity of the projectile's 
motion,~\cite{makoshi91,makoshikaji91} the Coulomb correlations on the projectile,~\cite{goldberg92,shao94,dutta01} 
and the collective electronic excitations of the surface.~\cite{vicentealvarez98} In addition, image shifts 
and level broadening due to image interactions as well as vibrations of the 
molecule may also be included in a theoretical description based on Green's functions. 

In contrast to Makoshi's work,~\cite{makoshi91,makoshikaji91} our approach is not restricted to time-local Auger 
self-energies and thus to the wide-band approximation. To overcome this limitation we solve the Dyson equation 
for the retarded Green's function by exponential resummation.~\cite{mahan00} Our approach is also not restricted 
to phenomenological Auger matrix elements. We work with the full matrix element, exploiting only the locality 
of the bound molecular states and the large distance of the molecule's turning point from the surface. 
%keeping 
%the dependencies on the single-electron quantum numbers alive. 
Although the final equations for the secondary 
electron emission coefficient and the spectrum of the Auger electron are highly complex they can be numerically 
evaluated within an interpolatory grid-based Monte-Carlo integration scheme. 

We specifically apply our approach to Auger de-excitation of \NitrogenDominantMetastableState\ on an 
aluminum or a tungsten surface. The metastable molecule is assumed to be in its vibrational ground state 
and the molecule's turning point is obtained from the surface potential applicable to 
\NitrogenDominantMetastableState\ on metallic surfaces.
%~\cite{katz95} 
For an aluminum surface we find the secondary electron emission coefficient due to Auger de-excitation 
(\ref{penning reaction}) to be one to two orders of magnitude smaller than the one deduced from 
\citeauthor{lorente99}'s~\cite{lorente99} theoretical study of the direct charge-transfer channel 
(\ref{direct reaction}) and for tungsten we find good quantitative agreement with 
an experimental estimate based on \citeauthor{stracke98}'s~\cite{stracke98} measurements. 
%Thus, however crude our model may be, it apparently captures
%the essential physics of Auger de-excitation of molecules at metallic surfaces.

The paper is structured in the following manner. In the next section we give more details about the effective 
model on which our analysis of Auger de-excitation is based. In Sect.~\ref{quantum kinetics} we employ  
the Keldysh technique to extract physical quantities from the model. Thereafter we describe in 
Sec.~\ref{numerical section} the numerical scheme we employed for the calculation of the secondary electron 
emission coefficient and the energy spectrum of the Auger electron. Results are presented in Sec.~\ref{results}. 
We conclude the paper in Sec.~\ref{conclusions} and complement it by two appendices. Appendix~\ref{wave functions} 
contains the explicit form of the wave functions we used in our calculations and Appendix~\ref{keldysh formalism} 
fixes the notations of the Keldysh formalism.

\section{Model\label{model}}

We investigate the de-excitation of a metastable nitrogen molecule impacting on a metallic surface with simultaneous 
release of an electron. The model we employ is an effective one that concentrates on the most important degrees of 
freedom and enables us to describe the system by a few parameters which are accessible through experiments or theoretical calculations. The 
primary goal will be to calculate the secondary electron emission coefficient~$\gamma_e$.

\begin{figure}[t]
        \includegraphics[width=.75\linewidth]{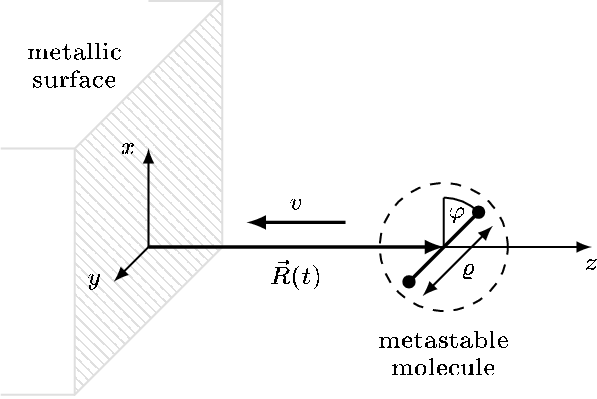}
        \caption{\label{surface impact}Schematic illustration of the collision geometry.}
\end{figure}
Focusing on the essentials of the process, we introduce from the start some simplifications and restrictions. First, 
we assume the metal surface to be planar, ideal, uncharged, and to stretch over the entire half space~${z<0}$. Furthermore, 
we consider only the dominant metastable state \NitrogenDominantMetastableState. In addition, we employ the trajectory 
approximation,~\cite{yoshimori86} that is, we decouple the translational motion of the molecule from the dynamics 
of the system and externally supply its trajectory. Finally, the molecule is assumed to impact the surface
under normal incidence with constant velocity $v$ and constant angle $\varphi$ of its axis to the surface.
Because of the translational symmetry of the solid surface in the ${x\text{-}y}$-plane, it is then sufficient
to consider only rotations of the molecule axis about one particular axis in this plane, for instance,
the $y$-axis (see Fig~\ref{surface impact}). 

We now cast the model assumptions into mathematical form, starting with the trajectory. Assuming the molecule to 
start moving at $t_0=-\infty$ and to hit the surface at $t=0$ the trajectory of its center of mass is 
\begin{equation}
	\vec{R}(t) = z_R(t) \, \vec{e}_z = \left( v |t| + z_0 \right) \vec{e}_z \,,
	\label{trajectory}
\end{equation}
where $z_0$ denotes the turning point.  The center of mass 
motion is classical. Hence, the turning point $z_0$ can be determined by considering the motion of the molecule 
in the  molecule-surface interaction potential $V_S(z)$ for given initial kinetic energy $\eps_{kin}$. Using 
a Morse-type potential,
%
%\begin{equation}
%	V_S(z) = d \left( 1 - e^{-a (z - z_e)} \right)^2 - d~,
%	\label{morse potential}
%\end{equation}
%
energy conservation gives for the position of the turning point
\begin{equation}
	z_0 = z_e - \frac{1}{a} \ln \left( 1 + \sqrt{1 + \frac{\eps_{kin}}{d}} \right)
	\label{turning point}
\end{equation}
with material parameters $d$, $a$, and $z_e$.~\cite{katz95} 
For metals 
\begin{equation}
	d = 0.738\,eV, \quad a = 640\,pm^{-1}, \quad z_e = 245\,pm.
	\label{morse parameters}
\end{equation}
According to the authors of Ref.~\onlinecite{katz95} these values are not very specific to the particular metal. 

\begin{figure}[t]
        \includegraphics[width=.98\linewidth]{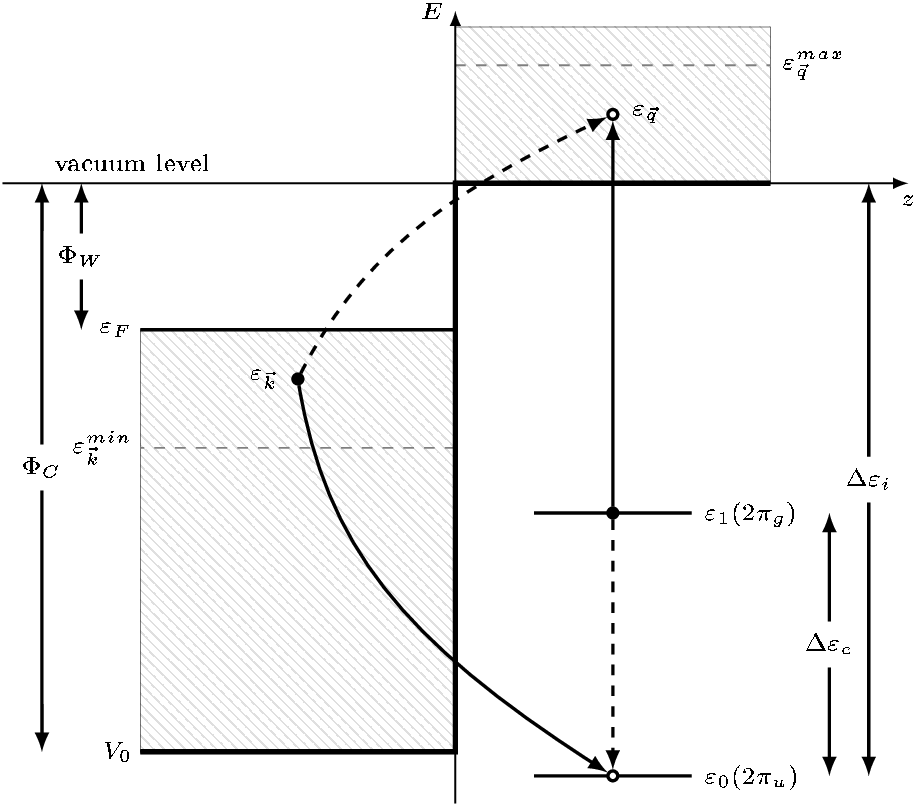}
        \caption{\label{energy scheme}Energy scheme of the simplified model showing Penning 
de-excitation (solid lines) and its exchange process (dashed lines). Also indicated are the 
classical energy cut-offs $\eps_{\vec{k}}^{min}$ and $\eps_{\vec{q}}^{max}$ which can be calculated from 
the energy balance $\eps_1 + \eps_{\vec{k}} = \eps_0 + \eps_{\vec{q}}$ that holds in the adiabatic limit.}
\end{figure}

To set-up, for a given trajectory, a Hamiltonian for a nitrogen molecule de-exciting at a metal surface we 
follow \citeauthor{Cazalilla98}~\cite{Cazalilla98} and distinguish between indirect de-excitation (Penning
process, solid lines in Fig.~\ref{energy scheme}), in which the electron is emitted from the molecule, and
direct de-excitation (exchange process, dashed lines in Fig.~\ref{energy scheme}), in which the electron 
is emitted from the surface. It must be stressed that although the asymptotic form of the wave function of the 
emitted electron is in both cases a plane wave, in the region relevant for the calculation of the matrix element, 
the Auger wave function for the Penning process resembles a single-electron continuum wave function of the 
nitrogen molecule whereas the Auger wave function for the exchange process is basically a continuum state of 
the solid with positive energy. In the following we will only consider the Penning process because our 
calculations showed that its matrix element is much larger than the matrix element associated with the exchange 
process, a consequence of the orthogonality of the bound molecular wave functions.

We construct the Hamiltonian by combining three different kinds of single-electron states to a 
single-electron basis: the single-electron states of the conduction band of the solid surface, which 
we approximate by the states corresponding to an electron trapped by a step potential of depth 
$\Phi_C$,~\cite{salmi92} the free single-electron states associated with the molecule's continuum
for which we use a two-center Coulomb wave,~\cite{joulakian96,weck02,yudin06} and effective single-electron 
states for the bound states of the molecule. To keep the description of the molecule as simple as possible 
we approximate the latter by a degenerate two-level system keeping, within the LCAO representation 
of the nitrogen molecule,~\cite{silbey04} only the $2\pi_u$ and the $2\pi_g$ molecular orbitals (MOs) 
which are the two MOs whose occupancies change during the de-excitation process. To construct the
two MOs we use moreover hydrogen-like wave functions with effective charges to mimic the 
Roothaan-Hartree-Fock wave functions of atomic
nitrogen.~\cite{clementi74} In the molecule's ground state \NitrogenGroundState~the $2\pi_u$ MO is fully 
occupied and the $2\pi_g$ MO is empty while in the excited state \NitrogenDominantMetastableState~the $2\pi_u$ MO 
contains a hole and the $2\pi_g$ MO is singly occupied. Both of these levels can carry four electrons and 
are degenerate in the electron spin ${s=\pm\frac{1}{2}}$ and the magnetic quantum number ${m=\pm1}$. Since 
the process we consider does not involve any spin flip, we ignore the spin. We can thus label the 
ground state of the two-level system and its excited state by $0m$ and $1m$, respectively, and denote the 
corresponding energies by $\eps_0$ and $\eps_1$. The states of the metal and the free states are 
labelled by $\vec{k}$ and $\vec{q}$, respectively. The mathematical expressions for the wave functions of
the single-electron states are given in Appendix A. 

The description of the electronic structure of the molecule-surface system is 
completed by aligning the single-electron states against each other and against the vacuum level by use 
of the metal's work function $\Phi_W$, the metal's conduction band depth $\Phi_C$, the molecule's ionization 
energy $\Delta\eps_i$, and the excitation energy of the molecule $\Delta\eps_e$. The metal states 
are of course occupied up to the Fermi level $\eps_F$. Our model is thus characterized by a few 
energy parameters, an effective charge, and a bond length (which enters the molecular wave functions 
of the diatomic molecule). 

The electronic structure of the simplified model is sketched in Fig.~\ref{energy scheme}, together with 
the transitions of the Penning de-excitation and its associated exchange process. Due 
to the symmetries of the molecular ground state (\NitrogenGroundStateTermSymbol) and the molecular excited state 
(\NitrogenDominantMetastableStateTermSymbol), only transitions with $\Delta m = 0$ are involved. They are driven 
by the Coulomb interaction between the excited electron in the $2\pi_g$ MO and an electron in the Fermi sea
of the metal (see, for instance, Refs.~\onlinecite{penn90,salmi92,bonini03,Cazalilla98}). The three electrons 
in the $2\pi_u$ MO 
act only as spectators and can thus be neglected. Assuming moreover the Fermi surface of the metal to be 
rigid, the de-excitation of \NitrogenDominantMetastableState~is basically a two-body scattering process, 
whose Hamiltonian, written in the single-electron basis described in the previous paragraph, is given by 
\begin{subequations}
\begin{align}
	H = \, & H_0 + H_1(t) \,, \\[0.5ex]
	\begin{split}
		H_0 = \, & \sum_{\vec k} \eps_{\vec k}^{\phdag} \, c_{\vec{k}}^{\dag} \, c_{\vec{k}}^{\phdag} + \sum_{\vec q} \eps_{\vec q}^{\phdag} \, c_{\vec{q}}^{\dag} \, c_{\vec{q}}^{\phdag} \\
		& + \sum_{m} \eps_0^{\phdag} \, c_{0m}^{\dag} \, c_{0m}^{\phdag} + \sum_{m} \eps_1^{\phdag} \, c_{1m}^{\dag} \, c_{1m}^{\phdag} \,,
	\end{split} \\
	\begin{split}
		H_1 (t) = \, & \sum_{\vec k, \vec q, m} \PenningME{m}(t) \, c_{0m}^{\dag} c_{\vec{k}}^{\phdag} \, c_{\vec{q}}^{\dag} \, c_{1m}^{\phdag} \,+\, H.c. \,,
	\end{split}
\end{align}\label{hamiltonian}%
\end{subequations}
%\begin{subequations}
%\begin{align}
%	H = \, & H_0 + H_1(t) \,, \\[0.5ex]
%	\begin{split}
%		H_0 = \, & \sum_{\vec k} \eps_{\vec k}^{\phdag} \, c_{\vec{k}}^{\dag} \, c_{\vec{k}}^{\phdag} + \sum_{\vec q} \eps_{\vec q}^{\phdag} \, c_{\vec{q}}^{\dag} \, c_{\vec{q}}^{\phdag} \\
%		& + \sum_{m} \eps_0^{\phdag} \, c_{0m}^{\dag} \, c_{0m}^{\phdag} + \sum_{m} \eps_1^{\phdag} \, c_{1m}^{\dag} \, c_{1m}^{\phdag} \,,
%	\end{split} \\
%	\begin{split}
%		H_1 (t) = \, & \sum_{\vec k, \vec q, m} \PenningME{m}(t) \, c_{0m}^{\dag} c_{\vec{k}}^{\phdag} \, c_{\vec{q}}^{\dag} \, c_{1m}^{\phdag} \\
%		& + \sum_{\vec k, \vec q, m} \PenningExchangeME{m}(t) \, c_{0m}^{\dag} c_{1m}^{\phdag} \, c_{\vec{q}}^{\dag} \, c_{\vec{k}}^{\phdag} \,+\, H.c. \,,
%	\end{split}
%\end{align}\label{hamiltonian}%
%\end{subequations}
%
where $H_0$ represents the Hamiltonian of the non-interacting system, and 
$H_1(t)$ contains the Penning de-excitation. 

The Auger matrix element $\PenningME{m}$ contains the time-dependence of the Hamiltonian, and thus carries the intrinsic 
non-equilibrium character of the system. In terms of the single-electron states given in Appendix A, it can be written as 
\begin{equation}
\begin{split}
	\PenningME{m}(t) = & \int \hspace{-0.3em} d\vec{r} \hspace{-0.1em} \int \hspace{-0.3em} d\vec{r}^\jvecprime \; \Psi_{0m}^{\conj} \bigl( \vec{r}_\varphi(t) \hspace{-0.1em} \bigr) \Psi_{\vec{k}}^{\phconj} \bigl( \vec{r}\, \bigr) \\
	& \times V_C \bigl( \abs{\vec{r} - \vec{r}^\jvecprime} \bigr) \Psi_{\vec{q}_\varphi}^{\conj} \bigl( \vec{r}_\varphi^\jvecprime(t) \hspace{-0.1em} \bigr) \Psi_{1m}^{\phconj} \bigl( \vec{r}_\varphi^\jvecprime(t) \hspace{-0.1em} \bigr)~,
\end{split} \label{penning matrix element}
\end{equation}
%\begin{subequations}
%\begin{align}
%	\begin{split}
%		\PenningME{m}(t) = & \int \hspace{-0.3em} d\vec{r} \hspace{-0.1em} \int \hspace{-0.3em} d\vec{r}^\jvecprime \; \Psi_{0m}^{\conj} \bigl( \vec{r}_\varphi(t) \hspace{-0.1em} \bigr) \Psi_{\vec{k}}^{\phconj} \bigl( \vec{r}\, \bigr) \\
%		& \times V_C \bigl( \abs{\vec{r} - \vec{r}^\jvecprime} \bigr) \Psi_{\vec{q}}^{\conj} \bigl( \vec{r}^\jvecprime \bigr) \Psi_{1m}^{\phconj} \bigl( \vec{r}_\varphi^\jvecprime(t) \hspace{-0.1em} \bigr)~,
%	\end{split} \label{penning matrix element} \\[1ex]
%	\begin{split}
%		\PenningExchangeME{m}(t) = & \int \hspace{-0.3em} d\vec{r} \hspace{-0.1em} \int \hspace{-0.3em} d\vec{r}^\jvecprime \; \Psi_{0m}^{\conj} \bigl( \vec{r}_\varphi(t) \hspace{-0.1em} \bigr) \Psi_{1m}^{\phconj} \bigl( \vec{r}_\varphi(t) \hspace{-0.1em} \bigr) \\
%		& \times V_C \bigl( \abs{\vec{r} - \vec{r}^\jvecprime} \bigr) \Psi_{\vec{q}}^{\conj} \bigl( \vec{r}^\jvecprime \bigr) \Psi_{\vec{k}}^{\phconj} \bigl( \vec{r}^\jvecprime \bigr)~,
%	\end{split} \label{exchange matrix element}
%\end{align}\label{matrix elements}%
%\end{subequations}
where $V_C$ represents the Coulomb potential and the subscript $\varphi$ denotes the associated vector as seen from 
the molecule's reference frame, which is centered about the molecule's center of mass $\vec{R}(t)$ and has its 
$z$-axis aligned along the molecule axis. The vectors 
$\vec{r}_\varphi^\jvecbracketprime(t)$ and $\vec{q}_\varphi$ are thus given by
\begin{subequations}
\begin{align}
	\vec{r}_\varphi^\jvecbracketprime(t) & = \hat\Omega(\varphi) \! \left( \vec{r}^\jvecbracketprime - \vec{R}(t) \!\right)~, \\
	\vec{q}_\varphi & = \hat\Omega(\varphi) \, \vec{q}~,
\end{align}
\end{subequations}
where the matrix $\hat\Omega(\varphi)$ describes the rotation around the $y$-axis
(see Fig. \ref{surface impact}).
%
%where the rotation matrix $\hat\Omega(\varphi)$ is defined as
%
%\begin{equation}
%	\hat\Omega(\varphi) = \begin{pmatrix}
%		\sin\varphi & 0 & -\cos\varphi \\
%		0 & 1 & \phantom{-}0 \\
%		\cos\varphi & 0 & \phantom{-}\sin\varphi
%	\end{pmatrix}~.
%\end{equation}
%
Due to the diatomicity of the nitrogen molecule the interaction matrix element depends on the 
orientation of the molecule with respect to the surface, that is, on the angle $\varphi$. For convenience 
we suppress however this dependence in our notation of the matrix element. Note, since we are 
considering the Penning process, the continuum states $\vec{q}$ are defined with respect to the
molecule's center of mass. Thus the associated wave function within Eq.~\eqref{penning matrix element} 
must depend on the molecular reference frame variables $\vec{q}_\varphi$ and $\vec{r}_\varphi^\jvecprime(t)$.

As already mentioned, we take the image interaction only for the emitted electron into account. It always 
feels the image potential and can thus only escape from the surface when its perpendicular energy is 
sufficiently high.~\cite{lorente99} Therefore, in the calculation of the secondary electron emission
coefficient and the spectrum of the Auger electron we will multiply the matrix element by a step function
\begin{equation}
	\Theta\!\left( \eps_{q_z} - \frac{e^2}{16 \pi \eps_0} \frac{1}{z_R(t) - z_i} \right)~,
	\label{surface transmission function}
\end{equation}
where $z_i$ is the position of the image plane. For aluminum and tungsten $z_i$ is given in 
Table~\ref{metal parameter table} and turns out to be closer to the surface than the turning point of 
the molecule. Equation~\eqref{surface transmission function} is thus well-defined along the 
trajectory of the molecule. It is sometimes referred to as surface transmission 
function,~\cite{lorente99} although for the case of Penning de-excitation this term is somewhat 
misleading, as the escaping electron is not emitted from the inside of the solid but from the molecule.

\section{Quantum kinetics\label{quantum kinetics}}

The model established in the previous section will now be treated using the Keldysh technique, a brief description 
of which is given in Appendix~\ref{keldysh formalism}. 

We start by calculating the unperturbed Green's functions 
$\mathcal{G}_{\alpha\beta}^{(0)}$, with $\alpha$ and $\beta$ representing any of the labels we used to characterize
the single-electron states of the system, $\vec{k}$, $\vec{q}$, $0m$, or $1m$. Since the time evolution of the 
free Green's functions is determined by $H_0$, they are diagonal, that is,
$\mathcal{G}_{\alpha\beta}^{(0)}\sim\delta_{\alpha\beta}$. For convenience we abbreviate the double subscript 
$\alpha\alpha$ by $\alpha$ in the following. Inserting the solutions of the interaction-free Heisenberg 
equations for the creation and annihilation operators appearing in model \eqref{hamiltonian}, the free propagators read
\begin{subequations}
\begin{align}
	i G_\alpha^{R\supzero}(t,t^\jprime) & = \Theta(t - t^\jprime) \, e^{\frac{i}{\hbar} \eps_\alpha (t - t^\jprime)}, \\
	i G_\alpha^{A\supzero}(t,t^\jprime) & = - \Theta(t^\jprime - t) \, e^{\frac{i}{\hbar} \eps_\alpha (t - t^\jprime)}, \\
	i G_\alpha^{K\supzero}(t,t^\jprime) & = \bigl[ 1 - 2 n_\alpha(t_0) \bigr] \, e^{\frac{i}{\hbar} \eps_\alpha (t - t^\jprime)},
\end{align}\label{free green's functions}%
\end{subequations}
where $n_\alpha(t_0)$ is the initial occupancy of the state $\alpha$ at $t_0 = -\infty$. 

In accordance with the 
model we introduced in the previous paragraph we assume the excited molecular level to be initially occupied 
with a single electron of magnetic quantum number $m=\mu$, that is, ${n_{1m}(t_0) = \delta_{m\mu}}$. The 
molecular ground state level, in contrast, is empty at $t_0$ because we neglect the spectator electrons. 
Hence, $n_{0m}(t_0) = 0$ for all $m$. The free electron states are also empty at $t_0$, implying 
$n_{\vec{q}}(t_0) = 0$, and the electronic states within the metal are initially filled up to the Fermi 
energy $\eps_F$, that is, $n_{\vec{k}}(t_0) = \Theta(\eps_F - \eps_{\vec{k}})$.

For the calculation of the full Green's functions $\mathcal{G}_{\alpha\beta}$ we need expressions for the 
self-energies $\Sigma_{\alpha\beta}$, which, in line with the work by \citeauthor{makoshi91},~\cite{makoshi91} 
we derive from a diagrammatic expansion up to second order in the Auger matrix element. This is justified 
because the Auger matrix element is in general very small. Because of the diagonality of the
unperturbed Green's functions the self-energies and thus the full Green's function are also diagonal.

\begin{figure}[t]
\includegraphics[width=.5\linewidth]{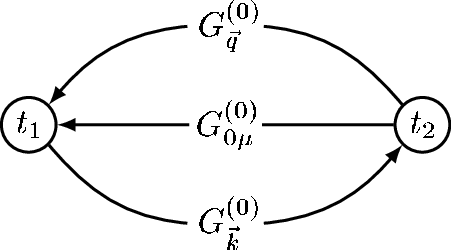}
\caption{\label{1 self-energy diagram}Diagrammatic representation of the self-energy $\Sigma_{1\mu}(t_1,t_2)$ of the
excited molecular state in second order perturbation theory.}
\end{figure}

We first investigate the excited molecular state. Figure~\ref{1 self-energy diagram} shows the only non-vanishing 
second-order self-energy diagram for $\Sigma_{1\mu}$. It can be evaluated to
\begin{subequations}
\begin{align}
	\Sigma_{1\mu}^R(t_1,t_2) = \, & \, \Theta(t_1 - t_2) \Sigma_{1m}^K(t_1,t_2), \label{sigma 1 retarded}\\
	\Sigma_{1\mu}^A(t_1,t_2) = \, & - \Theta(t_2 - t_1) \Sigma_{1m}^K(t_1,t_2), \\
	\begin{split}
		\Sigma_{1\mu}^K(t_1,t_2) = \, & - \frac{i}{\hbar^2} \sum_{\vec{q},\vec{k}} \left[ \PenningME{\mu}(t_1) \right]^\conj \PenningME{\mu}(t_2) \\
		& \times n_{\vec{k}}(t_0) \, e^{-\frac{i}{\hbar} \left( \eps_0 + \eps_q - \eps_k \right) (t_1 - t_2)}~.
	\end{split}\label{sigma 1 keldysh}
\end{align}\label{sigma 1}%
\end{subequations}
Using Eq.~\eqref{sigma 1 retarded}, the Dyson equation for the retarded Green's function
(Eq.~(\ref{gar dyson equation}) in Appendix B) can be solved iteratively. The result is
\begin{equation}
	G_{1\mu}^{R\phsupzero}(t,t^\jprime) = G_{\mu}^{R\supzero}(t,t^\jprime) W_\mu(t,t^\jprime)
	\label{gr series solution}
\end{equation}
with the infinite series
\begin{equation}
	W_\mu(t,t^\jprime) = \sum_{\nu=0}^\infty W_\mu^{(\nu)}(t,t^\jprime)
	\label{w series}
\end{equation}
whose individual terms, $W_\mu^{(\nu)}$, are given by 
%
%\begin{widetext}
%\begin{equation}
%	G_{1m}^{R\phsupzero}(t,t^\jprime) = G_{1m}^{R\supzero}(t,t^\jprime) \biggl[ 1 - \int_{t^\jprime}^{t_{\phantom{1}}} \hspace{-0.6em} dt_1 \hspace{-0.2em} \int_{t^\jprime}^{t_1} \hspace{-0.6em} dt_2 \; \Delta_{m}(t_1,t_2) + \int_{t^\jprime}^{t_{\phantom{1}}} \hspace{-0.6em} dt_1 \hspace{-0.2em} \int_{t^\jprime}^{t_1} \hspace{-0.6em} dt_2 \hspace{-0.2em} \int_{t^\jprime}^{t_2} \hspace{-0.6em} dt_3 \hspace{-0.2em} \int_{t^\jprime}^{t_3} \hspace{-0.6em} dt_4 \; \Delta_{m}(t_1,t_2) \, \Delta_{m}(t_3,t_4) - \,\dots \biggr],
%\end{equation}
%\begin{equation}
%	W_\mu^{(\nu)}(t_1,t_2) = (-1)^\nu \int_{t^\jprime}^{t_{\phantom{1}}} \hspace{-0.6em} dt_1 \hspace{-0.2em} \int_{t^\jprime}^{t_1} \hspace{-0.6em} dt_2 \hspace{-0.2em} \int_{t^\jprime}^{t_2} \hspace{-0.6em} dt_3 \hspace{-0.2em} \int_{t^\jprime}^{t_3} \hspace{-0.6em} dt_4 \, \dots \int_{t^\jprime}^{t_{2\nu-2}} \hspace{-1.9em} dt_{2\nu-1} \hspace{-0.2em} \int_{t^\jprime}^{t_{2\nu-1}} \hspace{-1.9em} dt_{2\nu} \; \Delta_\mu(t_1,t_2) \, \Delta_\mu(t_3,t_4) \dots \, \Delta_\mu(t_{2\nu-1},t_{2\nu})~,
%	\label{W series terms}
%\end{equation}
%\end{widetext}
\begin{equation}
\begin{split}
        W_\mu^{(\nu)}(t,t^\jprime) = & \; (-1)^\nu \int_{t^\jprime}^{t_{\phantom{1}}} \hspace{-0.6em} dt_1 \hspace{-0.2em} 
\int_{t^\jprime}^{t_1} \hspace{-0.6em} dt_2 \dots \hspace{-0.2em} \int_{t^\jprime}^{t_{2\nu-1}} \hspace{-1.9em} dt_{2\nu} \\[0.7ex]
&\times\Delta_\mu(t_1,t_2) \, \dots \, \Delta_\mu(t_{2\nu-1},t_{2\nu})~,
\end{split}\label{W series terms}
\end{equation}
where we introduced the quantity
\begin{equation}
        \Delta_\mu(t_1,t_2) = i \Sigma_{1\mu}^K(t_1,t_2) \, e^{\frac{i}{\hbar} \eps_1 (t_1 - t_2)}~,
        \label{delta 1m definition}
\end{equation}
which emerges from the self-energy terms of the iterated Dyson equation.

The infinite series~\eqref{w series} is exact but useless. To obtain an expression for the 
retarded Green's function which is amenable to further manipulations we employ the exponential resummation 
technique (see, for instance, Ref.~\onlinecite{mahan00}). For that purpose we introduce a new function 
$F_\mu(t,t^\jprime)$ and perform a perturbation expansion of $W_\mu$ and $F_\mu$ in terms of $\Delta_\mu$. 
Using the virtual expansion parameter $\lambda=1$ we write 
\begin{equation}
	\sum_{\nu=0}^\infty \lambda^\nu W_\mu^{(\nu)}(t,t^\jprime) = e^{F_\mu(t,t^\jprime)} = e^{\,\sum\limits_{\nu=1}^\infty \lambda^\nu F_\mu^{(\nu)}(t,t^\prime)}~.
	\label{w f series relation}
\end{equation}
Expanding the exponential in~\eqref{w f series relation} and then comparing the different orders of 
$\lambda$ leads to explicit expressions for the expansion coefficients $F_\mu^{(\nu)}$, the first few of 
which are
\begin{subequations}
\begin{align}
	F_\mu^{(1)} & = W_\mu^{(1)}, \\
	F_\mu^{(2)} & = W_\mu^{(2)} - \frac{1}{2} \left[ W_\mu^{(1)} \right]^2, \\
	F_\mu^{(3)} & = W_\mu^{(3)} - W_\mu^{(2)} W_\mu^{(1)} + \frac{1}{3} \left[ W_\mu^{(1)} \right]^3~, 
%	\dots \,. \nonumber
\end{align}\label{f coefficients}%
\end{subequations}
where, for convenience, we dropped the time arguments.

The retarded Green's function can now be conveniently written as
\begin{equation}
	G_{1\mu}^{R\phsupzero}(t,t^\jprime) = G_{1\mu}^{R\supzero}(t,t^\jprime) \, e^{F_\mu(t,t^\jprime)}
	\label{gr f form}
\end{equation}
with the function $F_\mu$ in the exponent given by the sum of the terms in~\eqref{f coefficients}. Using 
relation~\eqref{ga gr relation} together with Eq.~\eqref{gr f form} the advanced Green's function becomes 
\begin{equation}
	G_{1\mu}^{A\phsupzero}(t,t^\jprime) = G_{1\mu}^{A\supzero}(t,t^\jprime) \, 
        e^{\left[ F_\mu(t^\jprime,t) \right]^\conj}.
	\label{ga f form}
\end{equation}

To calculate the occupation of the excited molecular state we also need the Keldysh part of the Green's 
function defined in Eq.~\eqref{gk iterative solution} of Appendix B. Using the explicit form of the 
free Green's functions~\eqref{free green's functions} we first rewrite this equation into~\cite{makoshi91}
\begin{equation}
\begin{split}
	& G_\alpha^{K\phsupzero\!\!}(t,t^\jprime) = - i \bigl[ 1 - 2 n_\alpha(t_0) \bigr] G_\alpha^{R\phsupzero\!}(t,t_0) \, G_\alpha^{A\phsupzero\!}(t_0,t^\jprime) \\
	& \quad + \int_{t_0}^{t^{\phantom{\jprime}}} \hspace{-0.6em} dt_1 \hspace{-0.2em} \int_{t_0}^{t^\jprime} \hspace{-0.6em} dt_2 \; G_\alpha^{R\phsupzero\!}(t,t_1) \, \Sigma_\alpha^{K\phsupzero\!\!}(t_1,t_2) \, G_\alpha^{A\phsupzero\!}(t_2,t^\jprime)~,
\end{split}\label{gk makoshi}
\end{equation}
using the identity
\begin{equation}
\begin{split}
	G_\alpha^{K\supzero}(t,t^\jprime) = & - i \bigl[ 1 - 2 n_\alpha(t_0) \bigr] \\
	& \times G_\alpha^{R\supzero}(t,t_0) \, G_\alpha^{A\supzero}(t_0,t^\jprime)~,
\end{split}
\end{equation}
and the Dyson equation of the retarded and advanced Green's function~\eqref{gar dyson equation}. Note 
that Eq.~\eqref{gk makoshi} is not limited to the excited molecular level. It holds for all states~$\alpha$. 

Inserting~\eqref{sigma 1 keldysh}, \eqref{gr f form}, and \eqref{ga f form} into~\eqref{gk makoshi} we can now 
calculate the Keldysh part of the Green's function $G_{1\mu}^K(t,t^\jprime)$. Taking the latter at equal times 
$t=t^\jprime$ and utilizing Eq.~\eqref{green's occupancy} we finally obtain for the occupancy of the excited 
level at time $t$,
\begin{equation}
	n_{1\mu}(t) = e^{2 \jRe \left[ F_\mu(t,t_0) \right]}~,
	\label{n1 result}
\end{equation}
where $\jRe[\dots]$ denotes the real part. To lowest order in the interaction, that is, to lowest order in 
$\Delta_\mu$ the occupation of the molecular excited state is given by
\begin{equation}
	n_{1\mu}^{(0)}(t) = e^{2 \jRe \left[ F_\mu^{(1)}(t,t_0) \right]} = e^{- \hspace{-0.15em} 
        \int_{t_0}^t \hspace{-0.3em} dt_1 \hspace{-0.1em} \int_{t_0}^t \hspace{-0.3em} dt_2 \, \Delta_\mu(t_1,t_2)}~.
	\label{n1 result lowest order}
\end{equation}

We now turn to the free electron states, that is, the states which may get occupied by the 
electron released by the de-excitation of the molecule. A treatment of these states analogous to the excited 
state leads to the following expression for the occupation at time $t$
\begin{equation}
	n_{\vec{q}}(t) = 1 - e^{2 \jRe \left[ F_{\mu,\vec{q}}(t,t_0) \right]}~,
	\label{nq result equal footing}
\end{equation}
where $F_{\mu,\vec{q}}$ is defined the same way as $F_\mu$ but with $\Delta_\mu$ replaced by 
$\Delta_{\mu,\vec{q}}$. The latter is implicitly defined through the relation
\begin{equation}
	\Delta_\mu (t_1,t_2) = \sum_{\vec{q}} \Delta_{\mu,\vec{q}}(t_1,t_2)
	\label{delta m q definition}
\end{equation}
using Eqs.~\eqref{delta 1m definition} and \eqref{sigma 1 keldysh} for $\Delta_\mu (t_1,t_2)$.
 
Equation~\eqref{nq result equal footing} is not very useful, because it cannot easily be summed over 
$\vec{q}$, which is however needed to calculate the secondary electron emission coefficient $\gamma_e$.
Because of this obstacle we adopt the approach of Makoshi~\cite{makoshi91,makoshikaji91} and go back to 
%the Keldysh component of 
%the Green's function 
Eq.~\eqref{gk makoshi}, expand it for equal times $t=t^\jprime$ and $\alpha=\vec{q}$ up to first 
order in the self-energies, and insert the result into Eq.~\eqref{green's occupancy} which yields~\cite{makoshikaji91}
\begin{equation}
\begin{split}
	n_{\vec{q}}(t) = - i \int_{t_0}^{t^{\phantom{\jprime}}} \hspace{-0.6em} dt_1 \hspace{-0.2em} \int_{t_0}^{t^{\phantom{\jprime}}} \hspace{-0.6em} dt_2 \; & G_{\vec{q}}^{R\supzero}(t,t_1) \, \Sigma_{\vec{q}}^{K\phsupzero\!\!}(t_1,t_2) \\
	& \times G_{\vec{q}}^{A\supzero}(t_2,t^\jprime)~.
	\label{nq general equation}
\end{split}
\end{equation}

\begin{figure}
\includegraphics[width=.5\linewidth]{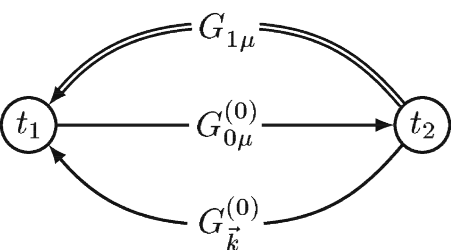}
\caption{\label{q self-energy diagram}Diagrammatic representation of the self-energy~$\Sigma_{\vec{q}}(t_1,t_2)$ of the emitted 
electron in second order perturbation theory. The dressed Green's function~$G_{1\mu}$ is indicated by a double line.}
\end{figure}

To account in Eq.~\eqref{nq general equation} for lifetime effects of the metastable molecule we follow once more
\citeauthor{makoshi91}~\cite{makoshi91} and employ in the calculation of the self-energy $\Sigma_{\vec{q}}^K$ 
the full ("dressed") Green's function of the excited state $G_{1\mu}^{+-}$ instead of the unperturbed one.
Up to second order (see Fig.~\ref{q self-energy diagram}) we obtain
\begin{equation}
\begin{split}
	& \Sigma_{\vec{q}}^K(t_1,t_2) = \frac{1}{\hbar^2} \sum_{\vec{k}} \PenningME{\mu}(t_1) \left[ \PenningME{\mu}(t_2) \right]^\conj \\
	& \quad \times G_{0\mu}^{-+(0)}(t_2,t_1) \, G_{\vec{k}}^{+-(0)}(t_1,t_2) \, G_{1\mu}^{+-\phsupzero\!}(t_1,t_2).
\end{split}\label{q keldysh self-energy}
\end{equation}

In order to proceed we need to calculate $G_{1\mu}^{+-}$. For that purpose, we first transform the corresponding 
component of the matrix Dyson equation \eqref{dyson equation} into
\begin{equation}
	G_{1\mu}^{+-\phsupzero\!\!} = G_{1\mu}^{+-\supzero} \! \left[ 1 + \Sigma_{1\mu}^{A\phsupzero\!\!} \, G_{1\mu}^{A\phsupzero} \right] \! + G_{1\mu}^{R\supzero} \, \Sigma_{1\mu}^{R\phsupzero\!\!} \, G_{1\mu}^{+-\phsupzero}~,
	\label{G1+- dyson equation}
\end{equation}
employing the fact that up to second perturbation order $\Sigma_{1m}^{++} \equiv \Sigma_{1\mu}^{R}$ and 
$\Sigma_{1\mu}^{--} \equiv - \Sigma_{1\mu}^{A}$. Inserting the free Green's functions~\eqref{free green's functions} 
as well as the self-energies~\eqref{sigma 1} and the full advanced Green's function~\eqref{ga f form} one can 
solve Eq.~\eqref{G1+- dyson equation} iteratively to obtain 
\begin{equation}
	G_{1\mu}^{+-\phsupzero\!\!}(t,t^\jprime) = G_{1\mu}^{+-\supzero}(t,t^\jprime) \, e^{F_\mu(t,t_0)} \, e^{\left[ F_\mu(t^\jprime,t_0) \right]^\conj}.
	\label{g1+- final}
\end{equation}
After inserting Eq.~\eqref{free green's functions} and \eqref{g1+- final} into Eq.~\eqref{q keldysh self-energy} we 
obtain for the Keldysh part of the self-energy
\begin{equation}
\begin{split}
	\Sigma_{\vec{q}}^K(t_1,t_2) = \, & \frac{i}{\hbar^2} \sum_{\vec{k}} \RenormalizedPenningME{\mu}(t_1) \! \left[ \RenormalizedPenningME{\mu}(t_2) \right]^\conj \\
	& \times n_{\vec{k}}(t_0) \, e^{\frac{i}{\hbar} \left( \eps_0 - \eps_k - \eps_1 \right) (t_1 - t_2)}
\end{split}\label{sigma q keldysh final}
\end{equation}
with 
\begin{equation}
	\RenormalizedPenningME{\mu}(t) = \PenningME{\mu}(t) \, e^{F_\mu(t,t_0)}
	\label{renormalized matrix element}
\end{equation}
the renormalized Auger matrix element.~\cite{makoshi91}

To compute finally the occupation of the $\vec{q}$-states, we insert Eq.~\eqref{sigma q keldysh final} into 
Eq.~\eqref{nq general equation} and obtain
\begin{equation}
	n_{\vec{q}}(t) = \int_{t_0}^{t^{\phantom{\jprime}}} \hspace{-0.6em} dt_1 \hspace{-0.2em} \int_{t_0}^{t^{\phantom{\jprime}}} \hspace{-0.6em} dt_2 \; \widetilde{\Delta}_{\mu,\vec{q}}(t_1,t_2)~,
	\label{nq result}
\end{equation}
where $\widetilde{\Delta}_{\mu,\vec{q}}$ is defined in Eq.~\eqref{delta m q definition} with the plain matrix 
elements replaced by the renormalized matrix elements leading to
\begin{equation}
	\widetilde{\Delta}_{\mu,\vec{q}}(t_1,t_2) = 
        \Delta_{\mu,\vec{q}}(t_1,t_2) \, e^{\left[ F_\mu(t_1,t_0) \right]^\conj} e^{F_\mu(t_2,t_0)}~.
\end{equation}

Equation~\eqref{nq result} represents the spectrum of the emitted electrons. The secondary electron
emission coefficient $\gamma_e$ can be calculated from Eq.~\eqref{nq result} 
by taking $t=\infty$ and summing over all possible $\vec{q}$,
\begin{equation}
	\gamma_e = \sum_{\vec{q}} n_{\vec{q}}(\infty) = \int_{t_0}^{\infty} \hspace{-0.6em} dt_1 \hspace{-0.2em} \int_{t_0}^{\infty} \hspace{-0.6em} dt_2 \; \widetilde{\Delta}_\mu(t_1,t_2)
	\label{gamma e result original}
\end{equation}
with $\widetilde{\Delta}_\mu$ defined by
\begin{equation}
	\widetilde{\Delta}_\mu(t_1,t_2) = \sum_{\vec{q}} \widetilde{\Delta}_{\mu,\vec{q}}(t_1,t_2)~.
\end{equation}
Equation~\eqref{gamma e result original} can be rewritten to
\begin{equation}
	\gamma_e = 1 - e^{2 \jRe \left[ F_\mu(\infty,t_0) \right]}~.
	\label{gamma e result}
\end{equation}
The lowest two orders of Eq.~\eqref{gamma e result} in terms $\Delta_\mu$ read
\begin{subequations}
\begin{align}
	\gamma_e^{(0)} = & \int_{t_0}^{\infty} \hspace{-0.6em} dt_1 \hspace{-0.2em} \int_{t_0}^{\infty} \hspace{-0.6em} dt_2 \; \Delta_\mu(t_1,t_2)~,
	\label{gamma e result lowest order} \\
	\gamma_e^{(1)} = & 1 - e^{- \hspace{-0.15em} \int_{t_0}^t \hspace{-0.3em} dt_1 \hspace{-0.1em} \int_{t_0}^t \hspace{-0.3em} dt_2 \, \Delta_\mu(t_1,t_2)}~.
    \label{gamma e result first order}
\end{align}\label{gamma e result lowest orders}
\end{subequations}

%\begin{widetext}
%\begin{subequations}
%\begin{align}
%	\gamma_e^{(0)} = & \sum_{\vec{q}} n_{\vec{q}}^{(0)}(\infty) = \sum_{\vec{q}} \int_{t_0}^{\infty} \hspace{-0.6em} dt_1 \hspace{-0.2em} \int_{t_0}^{\infty} \hspace{-0.6em} dt_2 \; \Delta_{\mu,\vec{q}}(t_1,t_2)~,
%	\label{gamma e result lowest order} \\
%	\gamma_e^{(1)} = & \sum_{\vec{q}} n_{\vec{q}}^{(1)}(\infty) = \sum_{\vec{q}} \int_{t_0}^{\infty} \hspace{-0.6em} dt_1 \hspace{-0.2em} \int_{t_0}^{\infty} \hspace{-0.6em} dt_2 \; \Delta_{\mu,\vec{q}}(t_1,t_2) \, e^{- \hspace{-0.15em} \int_{t_0}^{t_1} \hspace{-0.3em} dt_3 \hspace{-0.1em} \int_{t_0}^{t_3} \hspace{-0.3em} dt_4 \, \left[ \Delta_\mu(t_3,t_4) \right]^\conj} e^{- \hspace{-0.15em} \int_{t_0}^{t_2} \hspace{-0.3em} dt_5 \hspace{-0.1em} \int_{t_0}^{t_5} \hspace{-0.3em} dt_6 \, \Delta_\mu(t_5,t_6)}~.
%	\label{gamma e result first order}
%\end{align}\label{gamma e result lowest orders}
%\end{subequations}
%\end{widetext}

The preceding calculation of the self-energies does not treat free and excited states on an equal footing. Only 
the self-energy for the free states is renormalized whereas the one for the excited state is not. As a result,
particle conservation is not strictly guaranteed when the corresponding occupation numbers are calculated. 
The same shortcoming holds for \citeauthor{makoshi91}'s~\cite{makoshi91} original approach. Our numerical results 
showed however that particle conservation is obeyed for all physically relevant situations, justifying 
the treatment of the self-energies a posteriori.

Let us finally remark that although, as far as the logic of our approach is concerned, we closely 
followed \citeauthor{makoshi91},~\cite{makoshi91} our results have wider applicability. In contrast 
to him we do not work with a real phenomenological Auger interaction, depending only on time, and 
do not employ the wide-band approximation for the free states which would lead to time-local self-energies. 
We are also not restricted to the lowest order expressions given in Eq.~\eqref{n1 result lowest order} and 
\eqref{gamma e result lowest orders}. In principle, we can calculate corrections to these expressions 
using the higher order expansion coefficients of $F_\mu$ given by Eq.~\eqref{f coefficients}.

\section{Numerical scheme \label{numerical section}}

Inspecting the main results of the previous section, Eq.~\eqref{n1 result} and \eqref{gamma e result}, we 
realize that these equations are highly complex. For instance, calculating $\Delta_\mu$ from 
Eq.~\eqref{delta 1m definition} requires summation of a matrix element product over all $\vec{k}$ and $\vec{q}$ 
vectors, which, letting the box size ${L\rightarrow\infty}$, equals a sixfold integral. The Auger matrix element 
itself, according to Eq.~\eqref{penning matrix element}, involves another six dimensional integration over 
$\vec{r}$ and $\vec{r}^\jvecprime$. Since in Eq.~\eqref{delta 1m definition} the matrix element occurs as a product 
at two different times, this makes a total of eighteen dimensions of integration. Calculating the level 
occupancies from Eq.~\eqref{n1 result} and \eqref{gamma e result} requires at least another two-dimensional 
integration over the time arguments of $\Delta_\mu$. Thus, employing these equations as they stand for numerical 
calculations is clearly out of reach if reasonable computing time and considerably small numerical errors are 
required. In this section we will therefore introduce an approximation, which makes those numerical calculations 
feasible.

To simplify the matrix element~\eqref{penning matrix element} we first utilize the particular form of the wave 
functions (see Appendix~\ref{wave functions}). Since the molecular wave functions $\Psi_{0m/1m}$ are localized 
on the molecule and the metal wave functions $\Psi_{\vec{k}}$ and the free wave functions $\Psi_{\vec{q}}$ are 
bounded everywhere (in the mathematical sense), the main contribution to the matrix 
element~\eqref{penning matrix element} will arise from 
points close to the actual molecule position. Noting further, that for the low kinetic energies we are interested 
in ($\eps_{kin} \leq 1\,eV$) the molecule's turning point lies far outside the surface ($z_0 \geq 4.35\,a_B$, 
see Eq.~\eqref{turning point} and the parameters~\eqref{morse parameters}). We can thus safely restrict the 
$\vec{r}$-integration in~\eqref{penning matrix element} to $z\ge 0$, that is, neglect the overlap of the
wave functions inside the solid. Using then the transformation
\begin{equation}
	\vec{r}^{\jvecbracketprime\!} = \Omega^\dag(\varphi) \, \vec{r}_{1(2)} + \vec{R}(t)~,
\end{equation}
along with the explicit form of the electronic wave functions (see Appendix~\ref{wave functions}) we arrive at
\begin{equation}
	\PenningME{m}(t) \approx C_V T_{k_z} e^{-\kappa_{k_z} z_R(t)} \PenningMEStub{m}~,
	\label{matrix element separation}
\end{equation}
with
\begin{widetext}
\begin{subequations}
\begin{align}
	C_V = & \; \frac{e^2}{4 \pi \eps_0} \frac{\left[ N_{\bar{Z}}^q \right]^2}{\left( 2 \pi \right)^\frac{5}{2} \kappa^2 \sqrt{N_{2\pi_u} N_{2\pi_g}}} ~, \\
	\begin{split}
		\PenningMEStub{m} = & \int \hspace{-0.3em} d\vec{r}_1 \hspace{-0.1em} \int \hspace{-0.3em} d\vec{r}_2 \; \Theta\!\left( z_{1\varphi} + z_0 \right) \frac{\varrho_1 \varrho_2 \, e^{i m \left( \varphi_2 - \varphi_1 \right)}}{\abs{\vec{r}_1 - \vec{r}_2}} \left( e^{-\abs{\vec{r}_{1+}}} + e^{-\abs{\vec{r}_{1-}}} \right) \left( e^{-\abs{\vec{r}_{2+}}} - e^{-\abs{\vec{r}_{2-}}} \right) \\
		& \times e^{i \left( k_x x_{1\varphi} + k_y y_{1\varphi} \right)} \, e^{- \kappa_{k_z} z_{1\varphi}} \, e^{-i \vec{q}_\varphi \cdot\, \vec{r}_2} \, C\!\left( \vec{q}_\varphi, \vec{r}_{2+}, \bar{Z} \right) C\!\left( \vec{q}_\varphi, \vec{r}_{2-}, \bar{Z} \right)~,
	\end{split}\label{Vmkq}
\end{align}
\end{subequations}
\end{widetext}
where $\varrho_{i}$ and $\varphi_{i}$ ($i=1,2$) are the usual cylindrical coordinates and we have introduced 
the abbreviations $\bar{Z}=Z_C/2$, $\vec{r}_{i\pm}=\vec{r}_{i} \pm (\delta/2) \vec{e}_z$ ($i=1,2)$, and 
$\vec{r}_{1\varphi}=\left( x_{1\varphi}, y_{1\varphi}, z_{1\varphi} \right)=\hat\Omega^\dag(\varphi) \, \vec{r}_1$.
Here $\delta$ is the bond length of the molecule. Note that in Eq.~\eqref{Vmkq} we have used $\kappa^{-1}$ as the unit of length. Moreover, we did approximate $\Theta\!\left(z_{1\varphi} + z_R(t)\hspace{-0.1em}\right)$ by $\Theta\!\left(z_{1\varphi} + z_0\right)$, which, as numerical tests confirmed, is a good approximation for the distant turning point positions we are considering.
\begin{table}[t]
\begin{ruledtabular}
\begin{tabular}{c|r|r|r|r|r}
        Material & $\Phi_W\,[eV]$ & $\Phi_C\,[eV]$ & Ref. & $z_i\,[a_B]$ & Ref. \\\hline
        Al & 4.25 & 16.5 & \onlinecite{lorente99} & 2.95 & \onlinecite{jennings88} \\
%        Cu & 4.4 & 11.4 & \onlinecite{ashcroft76} \\
        W & 4.5 & 10.9 & \onlinecite{hagstrum54} & 3.0 & \onlinecite{jennings88}
\end{tabular}
\end{ruledtabular}
\caption{Model parameters for aluminum and tungsten.\label{metal parameter table}}
\end{table}

To calculate the occupation numbers from the equations given in Sec.~\ref{quantum kinetics}, we need to compute 
the time integral of $\Delta_\mu(t_1, t_2)$. Since in Eq.~\eqref{matrix element separation} the time dependence 
has been separated we can insert the trajectory~\eqref{trajectory} and carry out the time integration 
analytically. For the lowest order equations it suffices to calculate the integral with equal upper boundaries, that 
is,
\begin{equation}
	\mathcal{I}(t) = \int_{t_0}^t \hspace{-0.6em} dt_1 \hspace{-0.2em} \int_{t_0}^t \hspace{-0.6em} dt_2 \; \Delta_\mu(t_1,t_2)
\end{equation}
Using spherical coordinates for the wave vectors $\vec{k}$ and $\vec{q}$ we obtain
\begin{widetext}
\begin{equation}
\begin{split}
	\mathcal{I}(t) = \; & \frac{\bar{Z}^2}{2 \left( 2 \pi \right)^{8} N_{2\pi_u} N_{2\pi_g}} \frac{e^4}{\hbar^2 v^2 k_V^2 \eps_0^2} \int_0^{k_F} \hspace{-0.3em} dk \hspace{-0.1em} \int_0^\pi \hspace{-0.3em} d\vartheta_k \hspace{-0.1em} \int_0^{2\pi} \hspace{-0.3em} d\varphi_k \hspace{-0.1em} \int_0^{q_c} \hspace{-0.3em} dq \hspace{-0.1em} \int_0^{\frac{\pi}{2}} \hspace{-0.3em} d\vartheta_q \hspace{-0.1em} \int_0^{2\pi} \hspace{-0.3em} d\varphi_q \\
	& \times \frac{k^4 \sin\!\left( \vartheta_k \right) \cos^2\!\left( \vartheta_k \right) \sin\!\left( \vartheta_q \right)}{\left( 1 - e^{-2 \pi \frac{\bar{Z}}{q}} \right)^2} \, e^{-2 \kappa_{k_z} z_0} \abs{\PenningMEStub{\mu}}^2 \abs{\GammaFunction{t}}^2~,
\end{split}\label{delta time integral}
\end{equation}
where
\begin{align}
	k_F = \sqrt{\frac{2 m_e \left( \eps_F - V_0 \right)}{\hbar^2}}~,~~
	k_V = \sqrt{\frac{2 m_e \abs{V_0}}{\hbar^2}} ~,
\end{align}
and
\begin{equation}
\begin{split}
	\GammaFunction{t} = \; & \Theta\!\left( -t \right) \frac{e^{\left( \kappa_{k_z} - i \Delta\eps_{kq} \right) \min\left( t, -t_{\vec{q}} \right)}}{\kappa_{k_z} - i \Delta\eps_{kq}} + \Theta\!\left( t \right) \Biggl( \frac{e^{\left( \kappa_{k_z} - i \Delta\eps_{kq} \right) \min\left( 0, -t_{\vec{q}} \right)}}{\kappa_{k_z} - i \Delta\eps_{kq}} \\
	& + \Theta\!\left( t - t_{\vec{q}} \right) \frac{e^{- \left( \kappa_{k_z} + i \Delta\eps_{kq} \right) \max\left( 0, t_{\vec{q}} \right)} - e^{-\left( \kappa_{k_z} + i \Delta\eps_{kq} \right) t}}{\kappa_{k_z} + i \Delta\eps_{kq}} \Biggr)~,
\end{split}
\end{equation}
\end{widetext}
with
\begin{subequations}
\begin{align}
	\Delta\eps_{kq} & = \frac{\eps_0 + \eps_q - \eps_1 - \eps_k}{\hbar \kappa v} ~, \\
	t_{\vec{q}} & = z_i - z_0 + \frac{m_e e^2}{8 \pi \eps_0 \hbar^2 \kappa} \frac{1}{q^2 \cos^2\!\left( \vartheta_q \right)} ~.
\end{align}
\end{subequations}
The term $t_{\vec{q}}$ emerges from the inclusion of the transfer function~\eqref{surface transmission function} and 
needs to be set to zero when considering the population of the excited molecular level.

Equation~\eqref{delta time integral} is still highly complex, as it consists of an integral over 12 variables. In 
order to compute this expression numerically we divide the calculation into two steps. First, we calculate 
$\PenningMEStub{\mu}$ on a discrete grid within the {$(\vec{k},\vec{q}\hspace{0.1em})$-space}. Afterwards, 
we calculate the wave vector integral in Eq.~\eqref{delta time integral} while using sexa-linear interpolation 
to obtain the intergrid-values of $\PenningMEStub{\mu}$. Because of this grid-based scheme we have to introduce 
a cut-off constant $q_c$ for the $q$-integration in Eq.~\eqref{delta time integral}, which we choose to be slightly 
larger than the classical cut-off $q_{max}$. This is however not an issue, since the spectrum of the emitted electron 
falls off very rapidly beyond the classical cut-off. The high-dimensional integrals which
occur in our formalism are then efficiently computable by means of Monte Carlo techniques. In particular,
we employed the VEGAS algorithm as implemented in the GNU Scientific Library.
\begin{table}[b]
\begin{ruledtabular}
\begin{tabular}{c|r|r}
        Parameter & Value & Reference \\\hline
        $\delta$ & $2.067\,a_B$ & \onlinecite{silbey04} \\ % 109.4\,pm
        $\eps_0$ & $-17.25\,eV$ & \onlinecite{kaldor84} \\
        $\eps_1$ & $-9.57\,eV$ & \onlinecite{kaldor84}
\end{tabular}
\end{ruledtabular}
\caption{Model parameters for nitrogen.\label{molecule parameter table}}
\end{table}

\section{Results\label{results}}

We now present numerical results. The parameter values used in our calculations are listed in 
Table~\ref{metal parameter table} and \ref{molecule parameter table}. In all of the 
calculations we fixed the turning point $z_0$ to the value given by Eq.~\eqref{turning point} for 
$50\,meV$, which amounts to approximately $4.42\,a_B$. Within the energy range $\eps_{kin}\leq1\,eV$ 
this approximation is uncritical, since the turning point as given by Eq.~\eqref{turning point} varies 
very weakly. We aim at calculating, respectively, the occupancies of the excited molecular state $n_1(t)$ 
and of the free electron states $n(t)=\sum_{\vec{q}}n_{\vec{q}}(t)$. Within our model 
the initial magnetic quantum number $\mu=\pm 1$ has no impact on the occupation numbers. For 
convenience we thus omit any $m$ subscripts in the following. We restrict the molecule's orientation to 
the two fundamentally distinct situations $\varphi=0$ (axis parallel to the surface) and 
$\varphi=\frac{\pi}{2}$ (axis perpendicular to the surface). Furthermore, if not stated otherwise, we 
consider an aluminum surface.

We start our analysis with the final occupancies, that is, $n_1(\infty)$ and 
$n(\infty)$ where, according to Eq. \eqref{gamma e result original}, 
the latter quantity is the secondary electron emission coefficient $\gamma_e$. For now we employ 
only the lowest order equations~\eqref{n1 result lowest order} and~\eqref{gamma e result lowest order}, 
which, concerning the emitted electron, means that we neglect the matrix element renormalization and 
thus the lifetime effects of the metastable molecule. 

\begin{figure}[t]
	\includegraphics[width=.98\linewidth]{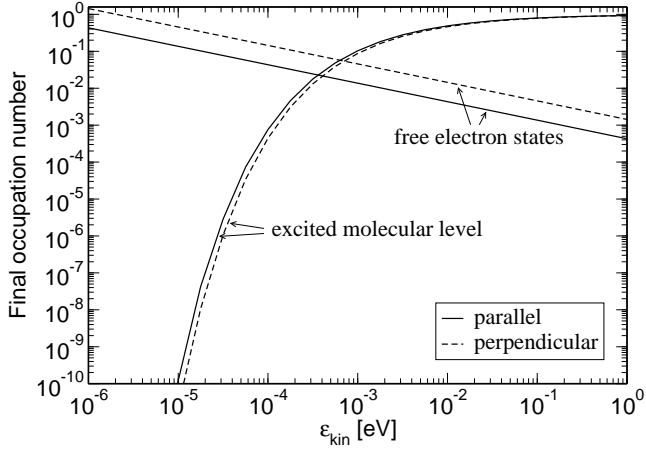}
	\caption{\label{n1 gamma (inf) fig}Final occupancies of the excited molecular level and 
of the free electron states (secondary electron emission coefficient) in parallel (solid lines) and 
perpendicular (dashed lines) molecule orientation for different kinetic energies $\eps_{kin}$ of 
the incident molecule. The curves were calculated using the zeroth order formulas, 
Eq.~\eqref{n1 result lowest order} and~\eqref{gamma e result lowest order}, respectively.}
\end{figure}
The final occupancies of the excited molecular level and of the free electron states
(secondary electron emission coefficient $\gamma_e$) are plotted in Fig.~\ref{n1 gamma (inf) fig} for different 
kinetic energies of the incident molecule. Obviously, the Penning process gets more efficient for lower kinetic 
energies, which is evident, because lower kinetic energies correspond to smaller molecule velocities and thus 
to larger interaction times of the solid-molecule system. Note, that for $\eps_{kin}\leq 10^{-6}\,eV$ the 
secondary electron emission coefficient gets larger than one although physical values for $\gamma_e$ should 
obviously be less than or equal to one. This unphysical peculiarity 
is a consequence of the negligence of the lifetime effect in the zeroth order 
formula for~$\gamma_e$, Eq.~\eqref{gamma e result lowest order}, and can be fixed by 
employing higher order terms of the full Eq.~\eqref{gamma e result} (see below).

From Fig.~\ref{n1 gamma (inf) fig} we see that for the two distinct orientations the Penning process is 
almost equally efficient in de-exciting the molecule. The number of emitted electrons, however, is about 
a factor three smaller when the molecular axis is parallel to the surface as compared to when it is 
perpendicular. Numerical tests showed that for the parallel case the electron is primarily emitted 
with a very small perpendicular energy whereas in the perpendicular case the perpendicular energy 
of the emitted electron is distributed more equally. Thus, in the perpendicular orientation the electron
has a higher probability to breach through the surface barrier originating from the image potential
leading to a larger value for $\gamma_e$.

\begin{figure}[t]
    \includegraphics[width=.98\linewidth]{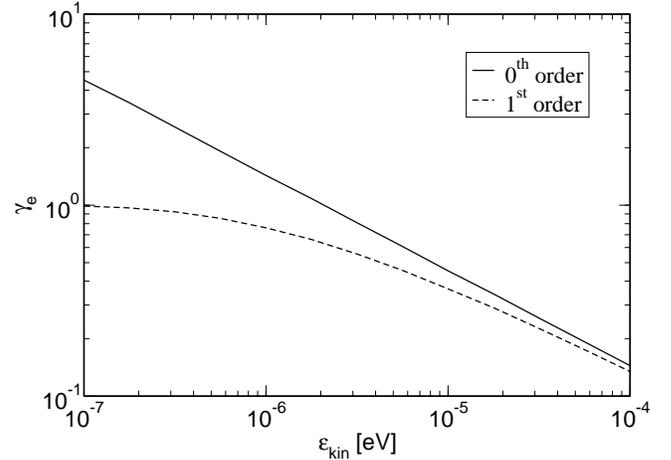}
        \caption{\label{gamma (inf) renormalized fig}Comparison of the secondary electron emission
coefficient $\gamma_e$ computed in zeroth order from Eq.~\eqref{gamma e result lowest order} (solid line)
and in first order from Eq.~\eqref{gamma e result first order} (dashed line). The molecular axis was aligned
perpendicular to the surface. Clearly, the first order correction heals the divergence of $\gamma_e$ for
$\eps_{kin} \rightarrow 0$ and ensures $\gamma_e \leq 1$ for all energies.}
\end{figure}

To fix the unphysical behavior of the secondary electron emission coefficient at low collision energies, 
the first order formula, Eq.~\eqref{gamma e result first order}, already suffices. It includes additional 
exponential factors which damp the integrand in the divergent region leading to 
$\gamma_e\rightarrow1$ for $\eps_{kin}\rightarrow0$. Figure~\ref{gamma (inf) renormalized fig} explicitly 
demonstrates this behavior. The region in which the zeroth order result for $\gamma_e$ exceeds unity 
corresponds to very low molecule velocities $\eps_{kin} \leq 10^{-6}\,eV$. These sub-thermal collision energies 
are rarely realized and are of lesser significance to our problem. In low-temperature plasmas, for instance, 
the systems we are primarily interested in, molecules have at least thermal collision energies. Only in beam 
experiments with extreme grazing incidence~\cite{winter02} may the collision energies be low enough to require 
the lifetime effect to be explicitly included in a theoretical analysis. For electron energies above 
$0.1~{\rm meV}$ the difference between the zeroth and first order formulas are vanishingly small indicating
that in this energy range Eq.~\eqref{gamma e result lowest order} is sufficient. 
%This is because the Auger 
%interaction is rather weak.

\begin{figure}[t]
        \includegraphics[width=.98\linewidth]{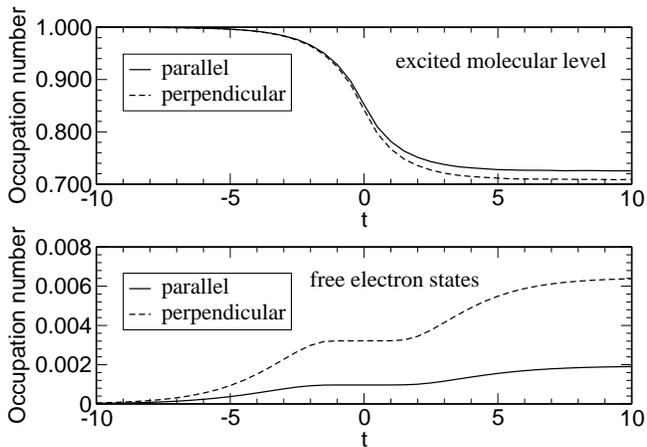}
    \caption{\label{n1 gamma (t) fig}Time evolution of the occupancies of the excited molecular level
(upper panel) and of the free electron states (lower panel) for parallel (solid lines) and perpendicular
(dashed lines) molecule orientation. The kinetic energy of the incident molecule was fixed to $50\,meV$.
Time is measured in units of $1/(\kappa v)$. A time difference of $\Delta t = 1$ thus corresponds
to the motion of the molecule over a distance of $\Delta z = a_B/2$.}
\end{figure}

Next, we investigate the time evolution of the occupancies. We fix the kinetic energy of the molecule
to $50\,meV$, which is about twice the thermal energy at room temperature. In addition, we employ the zeroth
order formulas. This is justified because the higher order corrections are small in the considered energy region. 
The results are plotted in Fig.~\ref{n1 gamma (t) fig}. Obviously, the occupancy of the excited molecular 
level (upper panel) changes significantly only in the range $\abs{t}\leq 5$, which, taking the turning point 
into account, equals maximum distances of the 
molecule's center of mass from the surface of roughly $7\,a_B$. 
%This implies that Penning de-excitation is 
%most efficient for small distances from the surface. 
As expected, within our model, the process is equally
effective in the incoming and outgoing branch of the trajectory. The time evolution of the occupancy of the 
free states (lower panel) is however distinctively different. It shows a plateau around $t=0$, that is, a stagnation 
of the probability to escape from the solid. This is a consequence of the image potential which almost completely 
traps an electron emitted at low surface distances where its perpendicular energy is too small to overcome
the image barrier encoded in the surface transmission function. 

\begin{figure}[t]
        \includegraphics[width=.98\linewidth]{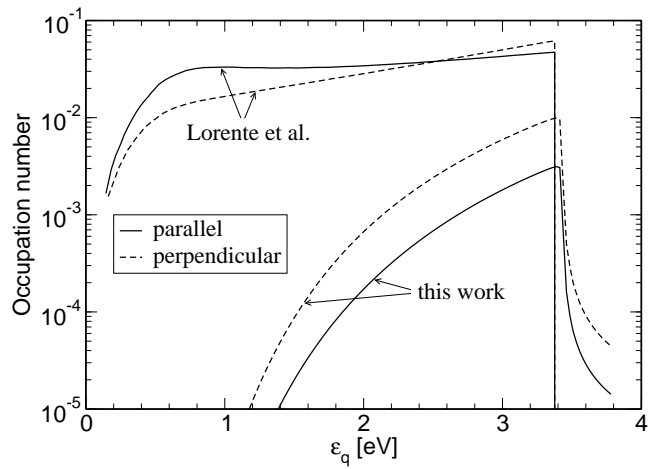}
        \caption{\label{nq fig}Energy spectrum $n_{\eps_{\vec{q}}}(\infty)=n_{\vec{q}}(\infty)$ of the 
Penning-emitted electron for parallel and perpendicular molecule orientation
(reaction (\ref{penning reaction}), this work). The kinetic energy of the 
incident molecule was fixed to $50\,meV$. For comparison we also show the results \citeauthor{lorente99}\cite{lorente99}
obtained for reaction (\ref{direct reaction}).}
\end{figure}

The energy distribution $n_{\eps_{\vec{q}}}$ of the emitted electron at $t=\infty$ is also of interest. This 
quantity is shown in Fig.~\ref{nq fig} for the two principal molecule orientations and a collision energy 
$\eps_{kin}=50\,meV$. The graphs for the two different orientations (denoted by \textquotedblleft this
work\textquotedblright) start at the origin and
monotonously increase until a cut-off energy is reached. The latter resembles the classical cut-off energy 
$\eps_{\vec{q}}^{max}$ (see Fig.~\ref{energy scheme}), implying that energy conservation is restored at the
end of the collision, as it should be. The low energy part of the spectrum is cut off due to the surface 
transmission function which allows electrons to escape from the surface only when their perpendicular 
energy is large enough. The spectrum for the perpendicular case takes on larger values, as was explained
in connection with the $\vec{q}$-integrated spectrum shown in Fig.~\ref{n1 gamma (inf) fig}.

For comparison, we also plotted in Fig.~\ref{nq fig} the spectra \citeauthor{lorente99}\cite{lorente99} 
obtained for an electron released due to charge-transfer and subsequent auto-detachment, reaction 
(\ref{direct reaction}), assuming $z_0\approx 5a_B$. For low energies the spectra due to Penning 
de-excitation (\ref{penning reaction}) are practically zero and thus significantly smaller than the ones 
corresponding to reaction (\ref{direct reaction}). 
In this energy range the resonant channel dominates. Close to the cut-off energy, however, the Auger
spectra are only one order of magnitude smaller than the spectra for the charge-transfer process 
indicating that in this energy range the two processes are indeed eye-to-eye competitors. To study in our model 
the competition between the two channels quantitatively requires however to dress-up the free Green's functions 
of the molecular electron by a hybridization self-energy and is thus beyond the scope of the 
present investigation. Integration of the energy spectra yields the secondary electron emission coefficients 
$\gamma_e$ listed in Table \ref{gamma table}. For aluminum the Penning process is thus one to two orders
of magnitude weaker than the resonant process.

\begin{table}[b]
\begin{ruledtabular}
\begin{tabular}{c|r|r}
	& $\gamma_e$ (parallel) & $\gamma_e$ (perpendicular) \\\hline
%    This work & ${2.83\cdot10^{-3}}$ & ${^{\phantom{1^1}}2.53\cdot10^{-3}}$ \\
    This work & ${1.93\cdot10^{-3}}$ & ${^{\phantom{1^1}}6.43\cdot10^{-3}}$ \\
	\citeauthor{lorente99}\cite{lorente99} & ${1.06\cdot10^{-1}}$ & ${^{\phantom{1^1}}8.98\cdot10^{-2}}$
\end{tabular}
\end{ruledtabular}
\caption{Comparison of the secondary electron emission coefficient $\gamma_e$ due to Penning de-excitation 
(this work) and resonant charge-transfer with subsequent auto-detachment (\citeauthor{lorente99}~\cite{lorente99}) 
for a \NitrogenDominantMetastableState\ molecule hitting an aluminum surface with a kinetic energy of $50\,meV$ 
and parallel or perpendicular orientation.\label{gamma table}}
\end{table}

\begin{figure}[t]
        \includegraphics[width=.98\linewidth]{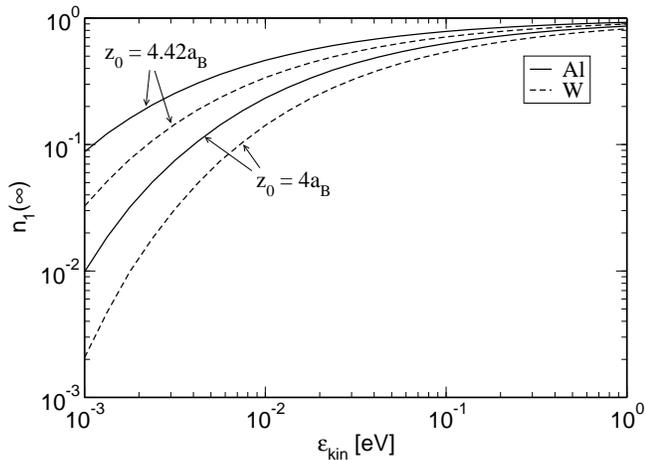}
        \caption{\label{gamma e turning point}Comparison of the final occupancy of the excited molecular 
level $n_1(\infty)$ for two different values of the turning point $z_0$ for aluminum (solid lines) and 
tungsten (dashed lines). The molecule's axis is perpendicular to the surface.}
\end{figure}

We now investigate the influence of the molecule's turning point on the de-excitation process. For aluminum 
and tungsten the occupation number of the excited molecular level calculated with the turning point given by 
Eq.~\eqref{turning point} for $\eps_{kin}=50\,meV$ is shown in Fig.~\ref{gamma e turning point} together with 
the data obtained for the turning point fixed to $z_0=4\,a_B$. For the latter the de-excitation probability increases 
drastically at low energies leading to significantly reduced occupancies of the excited molecular state. At 
large energies the difference in the turning points shows almost no influence. Quantitatively, for energies
below $1\,meV$, the occupancy of the excited molecular level for $z_0=4a_B$ is approximately one order of 
magnitude smaller as the one obtained for $z_0=4.42\,a_B$. This has however no influence for the occupancy 
of the free states (not shown in Fig.~\ref{gamma e turning point}) because an electron emitted closer to the 
surface has a lower perpendicular energy and can thus no longer breach through the surface potential arising 
from the image potential as we discussed before. 

Throughout our investigation we assumed that the Coulomb interaction, which drives the Penning de-excitation
of the molecule, is unscreened. In reality, however, the Coulomb interaction in the vicinity of a surface is 
screened due to the charge carriers of the solid. The strength of Penning de-excitation should thus be 
affected by screening. To estimate this effect it suffices to consider the statically screened Coulomb
potential $V_C^S(r)=V_C(r)e^{-\kappa_s r}$ with $\kappa_s$ the screening wave number at the surface. Little 
is known about this quantity except that it has to be smaller then the bulk screening wave number. Positron 
transmission and trapping experiments for various metallic films~\cite{neilson86} indicated, for instance, 
that the screening wave number near the surface is most probably a factor $0.6$ less than in the bulk. Taking 
this correction factor into account the screening wave number for an aluminum surface, for instance, 
is $1.2285 / \mathring{A}$. The screened Coulomb potential affects only the values of $\PenningMEStub{\mu}$, 
which enters quadratically into the lowest order formulas for the occupation numbers and reduces them 
by approximately 40\%. Thus, as expected, screening reduces the efficiency of the Coulomb-driven 
de-excitation channel, but it does not change its order of magnitude.

So far, we compared our results only with the theoretical results of \citeauthor{lorente99}\cite{lorente99}
who studied the charge-transfer channel (\ref{direct reaction}) for \NitrogenDominantMetastableState\ 
hitting an aluminum surface. To make contact to the experimental data obtained by 
\citeauthor{stracke98}~\cite{stracke98}, who also emphasized the direct charge-transfer reaction but 
nevertheless gave an estimate for $\gamma_e$ due to Penning de-excitation (\ref{penning reaction}), we 
also calculated $\gamma_e$ for tungsten using the $50\,meV$ turning point of the surface potential~\eqref{turning point} 
and a statically screened Coulomb interaction with $\kappa_s=0.6\,\kappa^W_b$ where $\kappa^W_b=1.7659/\mathring{A}$
is the screening wave number of bulk tungsten.~\cite{ashcroft76} We find $\gamma_e\approx 2 \cdot 10^{-3}$ 
which is rather close to the experimental estimate $\gamma^{\rm exp}_e|_{\rm Penning}\approx 10^{-4}-10^{-3}$.
Despite the crudeness of the effective model, which neglects, for instance, dangling bonds and 
surface states and the simplistic treatment of screening near a 
surface, our approach seems to capture even quantitatively the essential physics of Penning 
de-excitation. We attribute this to the rather large value of the molecule's turning point 
$z_0\approx 4.42a_B$ which partly immunizes Penning de-excitation against the details of the electronic 
structure in the immediate vicinity of the surface.

\section{Conclusions\label{conclusions}}

We investigated the release of secondary electrons due to Auger de-excitation of metastable nitrogen molecules 
at metallic surfaces using an effective model for the two active electrons involved 
in the process and the Keldysh formalism to calculate the occupation numbers of the relevant 
single-electron states as originally proposed by \citeauthor{makoshi91}.~\cite{makoshi91} In contrast to him, 
however, we are not restricted to time-local Auger self-energies and thus to the wide-band approximation
because we solve the Dyson equation for the retarded Green's function by exponential resummation. We also
employed the Auger matrix element obtained from a LCAO-type description of the molecule and an 
abrupt half-space-type description of the metal and not a phenomenological matrix element. The dependencies 
of the matrix element on the single-electron quantum numbers are retained in our calculation, as is 
the image interaction of the Auger electron with the surface and the distortion of the continuum of Auger
states in the vicinity of the surface due to the molecule which is accounted for by using for the Auger 
electron a continuum of two-center Coulomb waves.

The complexity of the final equations forced us to calculate and interpolate the Auger matrix element on a 
grid in a high-dimensional parameter space. This numerical approximation for the matrix element enabled us 
however to calculate the time evolution and the final values of the occupancies of the excited molecular level
and of the free electron states with standard Monte-Carlo integration routines without further approximations. 
Since the Auger interaction is
rather weak, we utilized only the lowest order formulas derived from the quantum kinetic theory. The lifetime 
correction introduced by~\citeauthor{makoshi91},~\cite{makoshi91} contained in higher order terms, was shown 
to be important only for very low kinetic energies of the molecule. 

We applied our approach to Auger de-excitation of \NitrogenDominantMetastableState\ on aluminum or tungsten. 
For an aluminum surface we showed that for realistic turning points of the molecule's trajectory, for instance, 
the one obtained from the molecular-surface potential for thermal kinetic energies, which is about $4.42a_B$, 
Auger de-excitation is one to two orders of magnitude less efficient in releasing an electron than the direct 
charge-transfer process investigated by Lorente and coworkers.~\cite{lorente99} For a tungsten surface our 
model produced for a turning point of $4.42a_B$ a secondary electron emission coefficient due to 
Auger de-excitation which agrees with the experimental estimates of Stracke and coworkers.~\cite{stracke98} 
%Auger de-excitation on a tungsten surface is only about one order of magnitude less efficient than 
%the direct charge-transfer process. It is thus strong enough to yield measurable effects.
%We also want to point out that for magnetic surfaces it is conceivable that the spin splitting works against the 
%resonance condition required for an efficient charge-transfer of an electron with the correct spin, that is, with 
%the spin of the missing electron in the ground state configuration of the molecule. In such a situation the Auger 
%process might even outperform the charge-transfer process. 

The effective model we used works with crude wave functions, but has the virtue to be parameterizable with a few 
easily obtainable energies. For the applications we have in mind, secondary electron emission in 
low-temperature gas discharges, where a great variety of different kinds of molecules and different kinds of 
wall materials occur, we consider this as a real advantage. 
With appropriate modifications the model can be applied to dielectric surfaces as well. We used the model only
for the investigation of Auger de-excitation. In combination with the Keldysh Green's function technique, however,
it can be also employed for the description of direct charge-transfer processes provided the energy shift and 
the broadening of the molecular levels due to the image interaction of the molecule with the surface and the 
self-energy corrections due to the Coulomb interaction between the excess electron and the
excited electron of the molecule are included.

\begin{acknowledgments}
	Johannes Marbach was funded by the International Max Planck Research School on Bounded Plasmas. In addition 
this work was supported by the Deutsche Forschungsgemeinschaft through the Transregional Collaborative Research 
Center SFB/TRR24.
\end{acknowledgments}

\appendix

\section{Wave functions\label{wave functions}}

The bound wave functions for the active molecular electron $\Psi_{0m/1m}$ are calculated from the linear combination
of atomic orbitals
\begin{equation}
	\Psi_{0m/1m} = \Psi_{2\pi_{u/g}^m} = \frac{\Psi_{2p_m}(1) \pm \Psi_{2p_m}(2)}{N_{2\pi_{u/g}}}
	\label{lcao wave functions}
\end{equation}
with $1$ and $2$ labeling the two distinct nitrogen atoms and $N_{2\pi_{u/g}}$ denoting normalization constants. The 
atomic nitrogen wave functions $\Psi_{2p_m}$ are approximated by using a hydrogen-like model with effective nucleus 
charge $Z_{\text{eff}}=4$. The resulting wave functions show excellent agreement with Roothaan-Hartree-Fock calculations 
for the nitrogen atom.~\cite{clementi74} The molecular wave functions~\eqref{lcao wave functions} are most conveniently 
expressed in cylindrical coordinates~${\left( \varrho, \varphi, z \right)}$ and possess the explicit form
\begin{equation}
\begin{split}
	& \Psi_{2\pi_{u/g}^m}(\varrho, \varphi, z) = \frac{1}{N_{2\pi_{u/g}}} \frac{- 2 m \kappa^{\frac{5}{2}}}{\sqrt{8 \pi}} \varrho e^{i m \varphi} \\[1.2ex]
	& \qquad \times \left( e^{-\kappa \sqrt{\varrho^2 + \left( z + \frac{\delta}{2} \right)^2}} \pm e^{-\kappa \sqrt{\varrho^2 + \left( z - \frac{\delta}{2} \right)^2}} \right),
\end{split}\label{molecular wave functions explicit}
\end{equation}
with $\kappa =2/a_B$ and normalization constants
\begin{equation}
\begin{split}
	N_{2\pi_{u/g}} = \; & 2 \pm 2 \kappa^5 \int_0^\infty d\varrho \int_{-\infty}^\infty dz \; \varrho^3 \\[1.2ex]
	& \times e^{-\kappa \sqrt{\varrho^2 + \left( z + \frac{\delta}{2} \right)^2}} e^{-\kappa \sqrt{\varrho^2 + \left( z - \frac{\delta}{2} \right)^2}}
\end{split}
\end{equation}
which need to be calculated numerically. The order of the wave functions within \eqref{lcao wave functions},
that is, the labeling of the nitrogen atoms does not matter in our calculation.

The wave functions for the electrons inside the metal are calculated along the lines of Ref.~\onlinecite{salmi92} 
by solving the Schr\"odinger equation for an electron trapped by the step potential
\begin{equation}
	V(z) = \begin{cases}
		- |V_0| & z < 0\\
		0 & z \geq 0
	\end{cases}.
\end{equation}
The presence of the molecule is thus ignored, as far as the calculation of the metal wave functions 
is concerned. Using box normalization with box size $L$ we obtain
\begin{equation}
\begin{split}
	\Psi_{\vec{k}}(\vec{r}) = & \frac{1}{L \sqrt{L}} \, e^{i \left( k_x x + k_y y \right)} \Bigl\{ T_{k_z} e^{-\kappa_{k_z} z} \Theta(z) \\
	& + \left[ e^{i k_z z} + R_{k_z} e^{-i k_z z} \right] \Theta(-z) \Bigr\}~,
\end{split}\label{metal wave function explicitly}
\end{equation}
where the following wave vector dependent coefficients have been introduced
\begin{subequations}
\begin{align}
	R_{k_z} & = \frac{i k_z + \kappa_{k_z}}{i k_z - \kappa_{k_z}}~, \\
	T_{k_z} & = \frac{2 i k_z}{i k_z - \kappa_{k_z}}~, \\
	\kappa_{k_z} & = \sqrt{\frac{2 m_e}{\hbar^2} \abs{V_0} - k_z^2}~. \label{kappa kz}
\end{align}\label{metal wave function parameters}
\end{subequations}
The energy of an electron in the state $\vec{k}$ is given by
\begin{equation}
	\eps_{\vec{k}} = \frac{\hbar^2}{2 m_e} \left( k_x^2 + k_y^2 + k_z^2 \right) - \abs{V_0}~.
\end{equation}

We now turn to the free states involved in the Penning process. As discussed in the main 
text we include here the effect the molecule has because the Auger electron originates from the 
molecule. The continuum of free states resembles thus not the continuum of the solid but the single-electron 
continuum of the molecule which we approximate by a two-center Coulomb (TCC) wave.~\cite{joulakian96}
In the past it has been successfully used to model electron-impact ionization and photo-ionization of 
$H_2$ and $H_2^+$ molecules.~\cite{weck02,yudin06} The TCC wave is an approximate solution of 
Schr\"odinger's equation for an unbound electron moving in the field of two fixed centers.
Employing $a_B$ as the unit of length it reads
\begin{equation}
	\Psi_{\vec{q}}(\vec{r}) = \frac{e^{i \vec{q} \cdot \vec{r}}}{\left( 2 \pi \right)^{\frac{3}{2}}} \, N_{Z_1}^q C\!\left( \vec{q}, \vec{r_1}, Z_1 \right) N_{Z_2}^q C\!\left( \vec{q}, \vec{r_2}, Z_2 \right)~,
	\label{tcc wave function}
\end{equation}
with 
%the normalization constants $N$ and the continuum factors $C$ given by
%
\begin{subequations}
\begin{align}
	N_Z^q & = e^{\frac{\pi}{2} \frac{Z}{q}} \,\, \Gamma\!\left( 1 + i \frac{Z}{q} \right) ~, \\[1.2ex]
	C\!\left( \vec{q}, \vec{r}, Z \right) & = M\!\left( -i \frac{Z}{q}, 1, -i \left[ q r + \vec{q} \cdot \vec{r} \, \right] \right)~,
\end{align}
\end{subequations}
where $M$ is the confluent hypergeometric function of the first kind. The two vectors $\vec{r}_1$ and $\vec{r}_2$ denote,
respectively, the position of the electron as seen from nucleus 1 and 2 and $Z_1$ and $Z_2$ are effective
charge numbers. As proposed in Ref.~\onlinecite{weck02} we account for the partial screening of the nuclei by the passive 
electrons and choose ${Z_1 = Z_2 = Z_C = \frac{1}{2}}$.

\section{Keldysh formalism\label{keldysh formalism}}

To fix our notation we give a brief description of the Keldysh formalism. For a more complete survey of the 
topic we refer the reader to Refs.~\onlinecite{keldysh65, blandin76, danielewicz84, rammer86}.

Consider a fermionic system with a Hamiltonian
\begin{equation}
	H(t) = H_0 + H_1(t) \,,
	\label{general time dependent hamiltonian}
\end{equation}
where $H_1(t)$ represents a time-dependent perturbation of the non-interacting system $H_0$. Due to the time-dependence 
of the Hamiltonian $H(t)$ we are faced with a non-equilibrium situation. 

One way to treat systems with a time-dependent Hamiltonian is the non-equilibrium Green's 
function technique introduced by Keldysh.~\cite{keldysh65} The key feature of the technique is a 
time contour $\mathcal{C}$ in the complex plane running from $-\infty$ to $\infty$ and then back 
again to $-\infty$. All quantities of the usual Green's function technique are then defined on this 
complex time path.
%(rather than on the real axis). 

Of particular importance is the contour-ordered Green's function 
$\mathcal{G}_{\alpha\beta}(t,t^\jprime)$, describing the propagation from a state $\beta$ at time $t^\jprime$ 
to a state $\alpha$ at time $t$. It is defined by 
\begin{equation}
	i \mathcal{G}_{\alpha\beta}(t,t^\jprime) = \left\langle T_{\mathcal{C}} \! \left[ \Psi_{\alpha}^\phdag(t) \, \Psi_{\beta}^{\dag}(t^\jprime) \right] \! \right\rangle_{\hspace{-0.3em}H}~,
	\label{green's function general}
\end{equation}
where $T_\mathcal{C}$ specifies the chronological time-ordering operator on the contour, 
$\Psi$ and $\Psi^\dag$ represent the usual field operators, and 
$\left\langle\dots\hspace{0pt}\right\rangle_{\hspace{-0.15em}H}$ denotes the averaging with respect to an 
arbitrary state of the full dynamical system~\eqref{general time dependent hamiltonian}. 

Employing the interaction picture Eq.~\eqref{green's function general} can be transformed to
\begin{equation}
	i \mathcal{G}_{\alpha\beta}(t,t^\jprime) = \left\langle T_{\mathcal{C}} \! \left[ \widetilde\Psi_{\alpha}^\phdag(t) \, \widetilde\Psi_{\beta}^{\dag}(t^\jprime) \, S_{\mathcal{C}} \right] \! \right\rangle_{\hspace{-0.3em}H_0}
	\label{green's function interaction picture}
\end{equation}
with the contour scattering operator $S_\mathcal{C}$ defined by
\begin{equation}
	S_{\mathcal{C}} = T_\mathcal{C} \, e^{-\frac{i}{\hbar} \int_\mathcal{C} \hspace{-0.15em} dt \, 
         \widetilde{H}_{1}(t)}~.
	\label{contour scattering operator}
\end{equation}
The tilde in Eq.~\eqref{green's function interaction picture} 
and~\eqref{contour scattering operator} characterizes the corresponding quantity in the interaction picture. 
Equation~\eqref{green's function interaction picture} is suitable for performing the usual perturbation 
expansion in terms of $\widetilde{H}_{1}(t)$, the only difference being that all time integrals need to be taken 
over the time contour $\mathcal{C}$ instead of the real time axis.

%\begin{figure}
%	\includegraphics[width=.8\linewidth]{contour.pdf}
%        \includegraphics[width=.8\linewidth]{Fig11.pdf}
%	\caption{\label{contour}Time contour in the complex plane indicating the increasing ($+$) 
%         and the decreasing ($-$) branch, as well as the turning point $\tau$ and a possible position of 
%       the time arguments $t$ and $t^\jprime$.}
%\end{figure}

Inspecting the definition of the contour ordered Green's function~\eqref{green's function general} and the possible location of the two time arguments $t$ and $t^\jprime$ on either the increasing ($+$) or the decreasing ($-$) branch of the contour we can decompose $\mathcal{G}_{\alpha\beta}(t,t^\jprime)$ into the four analytical pieces
\begin{subequations}
\begin{align}
	i G_{\alpha\beta}^{++}(t,t^\jprime) & = \left\langle T_{c} \! \left[ \Psi_{\alpha}^\phdag(t) \, \Psi_{\beta}^{\dag}(t^\jprime) \right] \! \right\rangle_{\hspace{-0.3em}H}~, \\
	i G_{\alpha\beta}^{+-}(t,t^\jprime) & = - \left\langle \Psi_{\beta}^{\dag}(t^\jprime) \, \Psi_{\alpha}^\phdag(t) \right\rangle_{\hspace{-0.3em}H}~, \label{G+- definition}\\
	i G_{\alpha\beta}^{-+}(t,t^\jprime) & = \left\langle \Psi_{\alpha}^\phdag(t) \, \Psi_{\beta}^{\dag}(t^\jprime) \right\rangle_{\hspace{-0.3em}H}~, \\
	i G_{\alpha\beta}^{--}(t,t^\jprime) & = \left\langle T_{a} \! \left[ \Psi_{\alpha}^\phdag(t) \, \Psi_{\beta}^{\dag}(t^\jprime) \right] \! \right\rangle_{\hspace{-0.3em}H}~,
\end{align}\label{green's function decomposition}%
\end{subequations}
where $T_c$ and $T_a$ denote the chronological and anti-chronological time ordering operator on the real time axis, 
respectively. Equation~\eqref{green's function decomposition} can be also expressed in matrix notation,
\begin{equation}
	\mathcal{G}_{\alpha\beta} = \begin{pmatrix}
		G_{\alpha\beta}^{++} & G_{\alpha\beta}^{+-} \\
		G_{\alpha\beta}^{-+} & G_{\alpha\beta}^{--}
	\end{pmatrix}~,
	\label{green's function matrix}
\end{equation}
where the time arguments are omitted for convenience. 

The time evolution of the Green's function~\eqref{green's function matrix} is 
governed by the Dyson equation
\begin{equation}
	\mathcal{G}_{\alpha\beta}^\phsupzero = \mathcal{G}_{\alpha\beta}^\supzero + \mathcal{G}_{\alpha\delta}^\supzero \, \Sigma_{\delta\gamma}^\phsupzero \, \mathcal{G}_{\gamma\beta}^\phsupzero \;,
	\label{dyson equation}
\end{equation}
where the summation over internal indices and integration over internal times is implicitly assumed. 
Equation~\eqref{dyson equation} involves the unperturbed Green's function (indicated by a $(0)$ superscript) 
and the self-energy $\Sigma_{\delta\gamma}$. The latter is defined on the contour $\mathcal{C}$ as well and thus 
possesses a matrix representation similar to Eq.~\eqref{green's function matrix}. Hence, the 
Dyson equation~\eqref{dyson equation} is a matrix equation. 

The structure of this equation can be simplified when it is noted that the set~\eqref{green's function decomposition} 
is linearly dependent. Applying the unitary transformation
\begin{equation}
	U = \frac{1}{\sqrt{2}} \left( \begin{array}{rr}
		1 & -1 \\
		1 & 1
	\end{array} \right)
	\label{green's transformation}
\end{equation}
to the Dyson equation~\eqref{dyson equation} the Green's function and the self-energy turn into
\begin{subequations}
\begin{align}
	\widehat{\mathcal{G}}_{\alpha\beta} & = U \, \mathcal{G}_{\alpha\beta} \, U^\dag =
	\begin{pmatrix}
		0 & G_{\alpha\beta}^A \\[1ex]
		G_{\alpha\beta}^R & G_{\alpha\beta}^{K}
	\end{pmatrix}, \label{G transfomed}\\[1ex]
	\widehat{\Sigma}_{\alpha\beta} & = U \, \Sigma_{\alpha\beta} \, U^\dag = \begin{pmatrix}
		\Sigma_{\alpha\beta}^{K} & \Sigma_{\alpha\beta}^R \\[1ex]
		\Sigma_{\alpha\beta}^A & 0
	\end{pmatrix}. \label{Sigma transfomed}
\end{align}
\end{subequations}
The superscripts $A$, $R$ and $K$ denote, respectively, the advanced, retarded and Keldysh part of the corresponding 
quantity. For the retarded and advanced Green's functions holds 
\begin{equation}
	G_{\alpha\beta}^A(t_1,t_2) = \left[ G_{\alpha\beta}^R(t_2,t_1) \right]^\conj.
	\label{ga gr relation}
\end{equation}

Carrying out the matrix multiplication in the transformed Dyson equation one obtains the set of 
equations that determines the different parts of the Green's function, 
\begin{subequations}
\begin{align}
	G_{\alpha\beta}^{A/R} = & \; G_{\alpha\beta}^{A/R\supzero} + G_{\alpha\delta}^{A/R\supzero} \, \Sigma_{\delta\gamma}^{A/R} \, G_{\gamma\beta}^{A/R} \;, 
	\label{gar dyson equation} \\
\begin{split}
	G_{\alpha\beta}^{K\phsupzero} = & \; G_{\alpha\beta}^{K\supzero} + G_{\alpha\delta}^{K\supzero} \, \Sigma_{\delta\gamma}^{A\phsupzero} \, G_{\gamma\beta}^{A\phsupzero} \\
	& + G_{\alpha\delta}^{R\supzero} \left[ \Sigma_{\delta\gamma}^{K\phsupzero} \, G_{\gamma\beta}^{A\phsupzero} + \Sigma_{\delta\gamma}^{R\phsupzero} \, G_{\gamma\beta}^{K\phsupzero} \right] .
\end{split}\label{gk dyson equation}
\end{align}
\end{subequations}

Equation~\eqref{gk dyson equation} can be solved iteratively to give the important relation
\begin{equation}
\begin{split}
	G_{\alpha\beta}^{K\phsupzero} = & \left[ \delta_{\alpha\gamma}^\phsupzero + G_{\alpha\delta}^{R\phsupzero} \, \Sigma_{\delta\gamma}^{R\phsupzero} \right] G_{\gamma\xi}^{K\supzero} \left[ \Sigma_{\xi\nu}^{A\phsupzero} \, G_{\nu\beta}^{A\phsupzero} + \delta_{\xi\beta}^\phsupzero \right] \\
	& + G_{\alpha\delta}^{R\phsupzero} \, \Sigma_{\delta\gamma}^{K\phsupzero} \, G_{\gamma\beta}^{A\phsupzero} \;.
\end{split}\label{gk iterative solution}
\end{equation}
The Keldysh part of the Green's function can thus be computed from Eq.~\eqref{gk iterative solution} once the 
advanced and retarded parts are known. The diagonal component $G_{\alpha\alpha}^K$ can then be used to 
calculate the occupation of the state $\alpha$ at arbitrary times
\begin{equation}
	n_\alpha (t) = \frac{1}{2} \left[ 1 - i G_{\alpha\alpha}^K(t,t) \right]~.
	\label{green's occupancy}
\end{equation}

%\bibliography{main}

\end{document}